\documentclass[fleqn,usenatbib]{mnras}

\usepackage{newtxtext,newtxmath}
\usepackage{amsmath}
\usepackage{graphicx}	
\usepackage[fleqn]{nccmath}

\usepackage[T1]{fontenc}

\DeclareRobustCommand{\VAN}[3]{#2}
\let\VANthebibliography\thebibliography
\def\thebibliography{\DeclareRobustCommand{\VAN}[3]{##3}\VANthebibliography}

\defcitealias{Gaensler2005}{G2005}

\renewcommand{\ne}{n_{\rm e}}
\newcommand{\RM}{\text{RM}}

\newcommand{\rad}{\text{rad}}

\newcommand{\cm}{\,{\rm cm}}    
\newcommand{\m}{\,{\rm m}}      
\newcommand{\pc}{\,{\rm pc}}     
\newcommand{\kpc}{\,{\rm kpc}}  

\newcommand{\Msun}{\,{\rm M}_{\odot}} 

\newcommand{\s}{\,{\rm s}}      
\newcommand{\Myr}{\,{\rm Myr}} 

\newcommand{\muG}{\,\mu{\rm G}} 

\newcommand{\Brms}{\,B_{\rm turb}}

\newcommand{\Hi}{H\,\textsc{i}}
\newcommand{\Hii}{H\,\textsc{ii}}
\newcommand{\Hei}{He\,\textsc{i}}
\newcommand{\Heii}{He\,\textsc{ii}}
\newcommand{\Heiii}{He\,\textsc{iii}}
\newcommand{\nH}{n_{\rm H}}

\newcommand{\erg}{\,{\rm erg}}  
\newcommand{\K}{\,{\rm K}}      

\newcommand{\gizmo}{\textsc{gizmo}}
\newcommand{\grackle}{\textsc{grackle}}
\newcommand{\slug}{\textsc{slug}}

\newcommand\rev[1]{#1}

\usepackage{graphicx}	
\usepackage{amsmath}	
\usepackage{hyperref}
\usepackage{orcidlink}

\newcommand{\aref}[1]{\hyperref[#1]{Appendix~\ref{#1}}}

\title[LMC Magnetised]{Wind of Change: Faraday Rotation in a Simulated Large Magellanic Cloud}

\author[Shah et al.]{
Hilay Shah,$^{\orcidlink{0000-0002-9136-6731}, 1}$\thanks{E-mail: Hilay.Shah@anu.edu.au}
Mark R.~Krumholz,$^{\orcidlink{0000-0003-3893-854X},1}$ and
N.\ M.\ McClure-Griffiths$^{\orcidlink{0000-0003-2730-957X},1}$
\\
$^{1}$Research School of Astronomy and Astrophysics, Australian National University, Canberra, ACT 2611, Australia\\
\\
}

\date{Accepted XXX. Received YYY; in original form ZZZ}

\pubyear{2015}

\begin{document}
\label{firstpage}
\pagerange{\pageref{firstpage}--\pageref{lastpage}}
\maketitle

\begin{abstract}
Magnetic fields significantly influence the structure of galaxies' interstellar media, but our understanding of magnetic field strengths and structures in external galaxies is severely limited. The Large Magellanic Cloud (LMC) offers a unique opportunity for improvement due to its proximity and large angular size, allowing for various detailed observations, particularly rich datasets of rotation measures and dispersion measures (RM and DM). However, interpreting these measurements is challenging due to the need for assumptions about the 3D structure for which we can only access \rev{line-of-sight} integrated quantities. To address this, we conduct a suite of high-resolution magnetohydrodynamic (MHD) simulations of the LMC, incorporating star formation, star-by-star feedback, and ram pressure stripping by the Milky Way's circumgalactic medium (CGM), experienced as a circumgalactic wind in the frame of the LMC. Synthetic observations of these simulations allow us to identify parameters that closely match observed RM and DM values. Our best model, which is an excellent match to the real LMC, yields magnetic field strengths of $\sim 1.4~\mu{\rm G}$ (ordered) and $\sim 1.6~\mu{\rm G}$ (turbulent). In this model, Milky Way CGM wind experienced by the LMC plays a critical role in shaping the RM data, with the bulk of the RM signal arising not from the LMC's plane, but from warm, $\sim 10^4$ K, gas in a Reynolds layer region $\sim 1$ kpc off the plane where relatively dense material stripped from the LMC is partially ionised by hard extragalactic radiation fields. This finding suggests that we should be cautious about generalising inferences from the LMC to other galaxies that may not be shaped by similar interactions.
\end{abstract}



\section{Introduction}
Magnetic fields permeate many astrophysical objects over a range of scales from planetary ($\sim 10^{-10}-10^{-8} \pc$) to intergalactic ($\sim 10^{6} \pc$). When these fields are strong, they have a broad range of physical effects that modify the evolution of the objects that host them. \rev{For example, they alter the way that fluids mix and cool, and suppress Kelvin-Helmholtz and Rayleigh-Taylor instabilities} \citep[e.g.,][]{Chandrasekhar1961, Ji2019, Das2023}. In the context of galaxy formation and evolution, simulations show that magnetic fields affect galactic properties including the $\rm H\textsc{~i}$ mass in the CGM, interstellar medium (ISM) disc sizes \citep{vandeVoort2021}, the extents of filamentary structures \citep{planck2016, federrath2016, Das2023}, star formation rates\rev{,} and initial stellar mass distributions \citep{Federrath2012, Krumholz2019}. Magnetic fields' effects on cosmic ray transport and feedback are also critical \citep{Salem2015, Seta2017, Farcy2022, Butsky2022}.

However, magnetic fields are notoriously difficult to observe. Polarised dust emission or absorption reveals the field direction, but not (at least not directly) its intensity. Field strengths can be measured using the Zeeman effect \citep{Crutcher12a}, but Zeeman splittings are generally too small to detect beyond the Milky Way. For extragalactic systems, perhaps our best available tool is the Faraday rotation measure (RM). \rev{We can use this tool in two ways: by observing either diffuse linearly polarised emission emitted by the extragalactic source itself \citep{Mao2012, Beck2020} or} bright and compact polarised background sources \rev{(most commonly active galactic nuclei, AGN, typically quasars or radio galaxies) that} emit linearly polarised synchrotron radiation \citep{Duric1988, KleinEA2015}. As this linearly polarised radiation passes through the magneto-ionic gas of a foreground galaxy, its \rev{intrinsic} polarisation angle $\psi_{0}$ rotates so that the observed polarisation angle becomes \rev{\citep{KleinEA2015, Ferriere2021}}
\begin{align}\label{eq:rm_angle}
        \psi_{\rm obs} = \psi_{0} + \RM~\lambda^2,
\end{align}
where $\lambda$ is the wavelength of the radiation and $\RM$ is the Faraday rotation measure, given by
\begin{align}\label{eq:rm_integral}
    \RM = 0.81~\int~\frac{n_e}{\cm^{-3}}~\frac{\vec{B}}{\muG}~\cdot~\frac{\vec{\mathrm{d}s}}{\pc}~\rad/\m^{2},
\end{align}
where $\ne$ is the thermal electron density, $\vec{B}$ is the component of the magnetic field along the line of sight \rev{(LOS)}, and $s$ measures position along the \rev{LOS} from \rev{the source to }our location \rev{\citep{Ferriere2021}}\footnote{\rev{Note that we drop the redshift dependence as all the sources of Faraday rotation we discuss in this paper are at $z\approx 0$.}}. By measuring how the polarisation angle of an observed source varies with wavelength, we can \rev{calculate} the value of RM along the LOS, which directly probes the mean field strength in the foreground gas. \rev{Because the background source method uniformly samples RMs across the target extragalactic source, while the diffuse emission method is biased by the locations of the emitting regions within it, the former is usually preferable for the purpose of determining the overall magnetic field structure despite the coarser sampling it offers.}

The Large Magellanic Cloud (LMC), one of the closest and largest Milky Way (MW) satellite galaxies, is an excellent target for extragalactic magnetic field studies both because it is an interesting system that is undergoing an interplay of complex interactions, and because it feasible to obtain large RM-based maps of its field structure. With regard to the first of these points, the LMC is currently subject to two major interactions \citep{Putman1998, Putman2003, StaveleySmith2003, Kim2003, Olsen2007, Nidever2013, Indu2015, Salem2015, Petersen2022}: ram pressure stripping and tidal interactions as a result of LMC's orbit through the MW circumgalactic medium (CGM), and tidal stripping due to the Small Magellanic Cloud (SMC) orbiting around it. These interactions significantly alter LMC's gas dynamics and presumably also its magnetic field structure. With regard to feasibility, observations of the LMC are facilitated by the fact that it resides just $\approx 50 \kpc$ from us \citep{Pietrzyski2019}, which makes its angular size as viewed from the Earth very large, $\rm \sim 10.75^\circ \times 9.17^\circ$, so that it covers a large number of background radio sources bright enough to permit RM measurements. Moreover, its location well off the MW plane (Galactic latitude $b\approx -33^\circ$) is fortuitous in that it lacks a significant MW foreground along the \rev{LOS}.

Given these attributes, it is not surprising that magnetic fields in the LMC have already been studied extensively. \citet[hereafter \citetalias{Gaensler2005}]{Gaensler2005} observed RMs along \rev{$\approx 100$} sightlines through the LMC and concluded that the fields show an axisymmetric spiral geometry, which they attributed to the action of a large-scale dynamo.
\citet{Livingston2024} followed the same observational methods and obtained similar qualitative results for the RM structure from a larger sample of sightlines. \citet{Mao2012} used the diffuse polarised synchrotron radiation arising from the LMC ISM to probe the magnetic fields in the side of the LMC halo nearer to us. Most of the $\RM$ values that \citeauthor{Mao2012} obtained are negative, meaning the magnetic fields are directed away from the observer\rev{. This is} consistent with the simple quadrupolar magnetic field structure expected for a large-scale dynamo  \citep{Parker1971}. 


While such an interpretation is plausible, it is by no means unambiguous, because it is far from trivial to infer magnetic fields from RM observations. As can be seen from \autoref{eq:rm_integral}, deriving the LOS magnetic field requires making assumptions about both the electron density and the path length through the magnetised medium. For instance, \citetalias{Gaensler2005}'s interpretation of the RM data as implying an axisymmetric field in the LMC disc relies on assuming a single electron column density for the entire LMC along with an idealised field geometry whereby halo magnetic fields are mirror-symmetric about the disc, leading their effects to cancel. \citet{Livingston2024} also assumed that RM measurements primarily probe disc fields, but adopted a slightly more complex model for position-dependent electron column densities. While adopting assumptions such as these is an understandable starting point given the difficulty of interpreting integrated LOS quantities, the LMC's complex environment \rev{nevertheless} renders them questionable. For example, the gas in the LMC halo should be strongly asymmetric due to tidal interactions and ram pressure stripping, and so the magnetic geometry is \rev{potentially asymmetric} -- and if there is no symmetry, then \rev{it is entirely possible} that RM measurements are primarily probing \rev{halo} fields rather than \rev{disc ones}. This complexity suggests a need to move past simple assumed geometries and instead combine realistic simulations with mock observations to facilitate more accurate inference of magnetic field structure.
 
With modern computers and simulation methods such simulations are possible, and indeed several authors have already simulated the LMC and its interactions with the MW and the SMC \citep[e.g.,][]{Mastropietro2005, Mastropietro2009, Besla2012, Salem2015, Pardy2018, Wang2019, Lucchini2020, bustard2020, Lucchini2021, Lucchini2024, JimnezArranz2024, Sheng2024}. These papers have examined topics including the role of tidal interactions in shaping the morphology of the LMC, the effects of ram pressure stripping on its \Hi\ gas disc, the orbital histories of the LMC-SMC system, features of the CGM surrounding the LMC\rev{, the effects of cosmic ray driven outflows in the presence of ram pressure stripping,} and the origin of gas in the Magellanic stream and bridge. However, none of these existing simulations have 
\rev{attempted to carry out systematic comparisons between simulations and the RM data in order to constrain the LMC's magnetic structure. Indeed, all of these simulations with the exception of \citeauthor{bustard2020} omitted magnetic fields altogether, and} it is entirely possible that \rev{this} omission leads to substantial inaccuracies in the simulation results, since there is ample evidence that magnetic fields have important dynamical effects in local galaxies \citep[e.g.,][]{Beck2009, Pakmor2013, Pakmor2017, Kim2019, Pillepich2019, Hopkins2019, vandeVoort2021, Wibking2023}. 

Here, we present the first-ever suite of high-resolution magnetohydrodynamic (MHD) simulations of the LMC and its interaction with the MW CGM. We post-process this suite to generate mock observables that permit direct comparison with measurements of the real LMC. In \autoref{sec:methods}, we outline our simulation setup and post-processing pipeline for generation of the mock observables. \autoref{sec:results} outlines our results and presents a detailed comparison between mock and real observations, which we use to identify which of the simulations in our suite best match reality. We in turn use this information to interpret several aspects of the RM data. 
Finally, in \autoref{sec:conclusions} we summarise our primary results.

\section{Methods}
\label{sec:methods}
Here we outline all the steps in our analysis: how we simulate the LMC (\autoref{sec:Simulation}), the initial conditions we use for those simulations (\autoref{sec:InitialConditions}), how we generate those mock observations from the simulations (\autoref{sec:MockObservables}), and the post-processing of the gas ionisation state required for the mock observations (\autoref{sec:PostProcess}).

\subsection{Simulation}
\label{sec:Simulation}
We use the \gizmo~code \citep{Hopkins2015std} to solve the equations of ideal magnetohydrodynamics (MHD) plus self-gravity for a fluid interacting with a population of collisionless star and dark matter particles\rev{. For} MHD, we use \gizmo's meshless finite mass (MFM) setting \citep{Hopkins2015mhd} with the constrained-gradient divergence cleaning method described by \citet{Hopkins2016divb}. Apart from MHD, we incorporate a time-dependent chemistry network with abundances of \Hi, \Hii, \Hei, \Heii, \Heiii, and free electrons\rev{. We also include} gas cooling using tabulated H, He, and metal cooling rates calculated with the photo-ionisation code \textsc{Cloudy} \citep{Ferland1998, Cloudy23}\rev{. Both of these components are }included as part of the \grackle~chemistry and radiative cooling library \citep{Smith2008, Smith2016}. \textsc{Grackle} includes a UV background radiation field based on the models of \citet{HM2012}, which for our simulations we evaluate at redshift $z=0$; it also includes an FUV background that provides a uniform photoelectric heating rate of $8.5 \times 10^{-26} \nH^{-1} \erg \cm^{-3} \s^{-1}$, where $\nH$ is the number density of H nucleons, in gas with temperatures $T < 2\times 10^4$ K, consistent with the estimates of \citet{WolfireEA2003} and \citet{Tasker&Bryan2008}. We use \grackle~rather than \gizmo's built-in cooling module because the former does not correctly produce a multi-phase atomic medium \citep{Wibking2023}, something that will be important for our purposes.

Our star formation model follows the approach described by \citet{Springel&Hernquist2003}\rev{. In this model,} gas elements with densities $\rho > \rho_\mathrm{SF}$ are probabilistically converted into star particles \citep[e.g.,][]{Katz1996}\rev{. The} probability per unit time \rev{is set }such that the expected star formation rate per unit volume for a fluid of density $\rho$ is $\dot{\rho}_* = \epsilon_\mathrm{ff} \rho / t_\mathrm{ff}$, where $t_\mathrm{ff} = \sqrt{3\pi/32 G \rho}$ is the gas free-fall time. Our settings for $\rho_\mathrm{SF}$ and $\epsilon_\mathrm{ff}$ mirror those described by \citet{Wibking2023} and \citet{Hu2023}. We set $\rho_\mathrm{SF}$ to a resolution-dependent value chosen so that star formation occurs only once the gas density exceeds our ability to resolve the collapse any further. Quantitatively, we set $\rho_\mathrm{SF}$ implicitly via the condition that $\Delta m = M_J(\rho_\mathrm{SF}, T_\mathrm{eq,SF})$, where $\Delta m$ is the simulation mass resolution, $M_J$ is the Jeans mass as a function of density and temperature, and $T_\mathrm{eq, SF}$ is the equilibrium temperature for gas of density $\rho_\mathrm{SF}$. We solve this equation numerically for our chosen heating and cooling rates and mass resolution, and report the resulting value of $\rho_\mathrm{SF}$ for each of our simulations below. For gas with density $\rho$ from $\rho_\mathrm{SF}$ to $100\rho_\mathrm{SF}$ we set $\epsilon_\mathrm{ff} = 0.01$, consistent with observations \citep[e.g.,][]{Krumholz07, Krumholz12, Krumholz19, Utomo18, Sun23}. In gas denser than $100\rho_\mathrm{SF}$ we artificially increase $\epsilon_\mathrm{ff}$ to 100\% so that gas is consumed very quickly; this limits the computational resources spent following high-density structures that evolve on short dynamical times. 

Our simulations include three channels of stellar feedback: stellar winds, supernovae, and photoionisation. We implement stellar wind and supernova feedback using the explicit mechanical feedback coupling algorithm described by \citet{Hopkins2014feedback} and \citet{Hopkins2018feedback}\rev{. This algorithm} ensures momentum and energy conservation as well as statistical isotropy\rev{. It} interpolates between treating SN feedback as a deposition of thermal energy and a deposition of mechanical momentum depending on whether the Sedov-Taylor radius is resolved. Our implementation of this algorithm is identical to that described in \citet{Wibking2023}. Our treatment of photoionisation uses a Str\"omgren volume method as re-implemented by \citet{Armillotta2019} based on the work of \citet{Hopkins2018photo}. Gas that is located within a Str\"omgren sphere is heated to a temperature of $10^4$ K.

Because we will carry out simulations with significantly higher mass resolution \rev{than} is typical in the cosmological applications for which \gizmo~is most commonly used, we do not use the standard \gizmo~method for computing the supernova, wind, and ionising photon output\rev{. The standard method} assumes stellar populations large enough to fully sample the stellar initial mass function (IMF). We instead use the stochastic star-by-star treatment outlined in \citet{Hu2023}. In this treatment when a star particle of mass $M_*$ forms, we stochastically draw a stellar population of that mass from a Chabrier IMF \citep{Chabrier2005} using the \slug~(Stochastically Lighting Up Galaxies) stellar population synthesis code \citep{daSilva2012, Krumholz2015}. We then run \slug~as a backend to the simulation, using it to track the evolution of the stellar population as it ages and return the instantaneous ionising luminosity and wind luminosity, and the timing of all supernovae, each of which is assumed to detonate with an energy budget of $10^{51}$ erg. We refer readers to \citeauthor{Hu2023} for a more detailed description of the method.

\subsection{Initial conditions}
\label{sec:InitialConditions}

All our simulations begin with gas, dark matter (DM) and old stellar\footnote{We use the term old stellar particles to indicate stars that are present at the start of a simulation, rather than forming during it. Old stellar particles differ from those formed in the course of the simulation in that they do not supply any feedback.} particles, whose masses are given in \autoref{tab:resolution_table} for our different resolutions. In all cases we set the initial temperature of all gas particles \rev{in the disc} to $10^4 \K$. \rev{Except} in very rarefied regions\rev{,} the gas cools from this temperature rapidly \rev{and collapses to form stars and inject supernovae feedback which aids in quickly attaining a gaseous and stellar steady-state \citep{Wibking2023}}. Similarly, we set the metallicity of all gas particles to $Z = 0.5Z_\odot$, the observed metallicity of the LMC \rev{\citep{Russell1992}}. We run all simulations for 500 Myr, roughly 5 rotation periods for an LMC-like galaxy, which we find is long enough for the structure of the galaxy to reach statistical steady state.

All other aspects of our initial conditions vary from one simulation to another. This variation is characterised by four parameters: the mass resolution $\Delta m$, the initial strengths of the smooth azimuthal and turbulent components of the magnetic field $B_0$ and $\Brms$, and whether we simulate an isolated galaxy or one that is traversing and subject to ram pressure stripping by the MW CGM. We list the full set of simulations we run in \autoref{tab:master_table}, and report resolution-dependent properties of the simulations in \autoref{tab:resolution_table}. For convenience, from this point forward we refer to simulations by names of the form \texttt{I-A-T-r} or \texttt{W-A-T-r}\rev{. In this naming convention,} \texttt{I} or \texttt{W} indicates whether the simulation uses an isolated (\texttt{I}) galaxy or one experiencing a wind (\texttt{W}) from the MW CGM, \texttt{A} and \texttt{T} give the strength of the initial smooth azimuthal and turbulent fields, respectively, in units of $\mu$G, and \texttt{R} is \texttt{L}, \texttt{M}, or \texttt{H}, indicating low ($\Delta m = 1000$ M$_\odot$ for gas), medium ($\Delta m = 250$ M$_\odot$), or high ($\Delta m = 100$ M$_\odot$) resolution. Thus for example \texttt{I-2-6-L} indicates a low-resolution simulation of an isolated galaxy with an initial magnetic field of 2 $\mu$G in a smooth azimuthal component and 6 $\mu$G in a turbulent component, while \texttt{W-0-12-M} indicates a medium-resolution simulation of an LMC experiencing a wind, and initialised with a 12 $\mu$G turbulent field and no smooth field. We describe the details of how we implement each of these aspects of the initial conditions below.

\begin{table}
\centering
\caption{Mass resolutions and star formation thresholds}
\begin{tabular}{ccccc}
\hline\hline
\\[-2ex]
Resolution \rev{(1)} & $\Delta m_{\rm gas}$ \rev{(2)} & $\Delta m_{\rm DM}$ \rev{(3)} & $\Delta m_*$ \rev{(4)} & $\rho_{\rm SF}/m_\mathrm{H}$ \rev{(5)} \\
& (M$_\odot$) & (M$_\odot$) & (M$_\odot$) & (cm$^{-3})$ 
\\[0.1ex]
\hline
\\[-2ex]
Low & $1000$ & $80682$ & $1136$ & $\phantom{0}194$\\
Medium & $\phantom{0}250$ & $20170$ & $\phantom{0}284$ & $1018$\\
High & $\phantom{0}100$ & $\phantom{0}8068$ & $\phantom{0}114$ & $2379$
\\ \hline\hline
\end{tabular}\\[1.5ex]
\footnotesize{Columns: (1)\rev{: Resolution -- resolution type}; (2)\rev{: $\Delta m_{\rm gas}$ -- gas particle mass}; (3)\rev{: $\Delta m_{\rm DM}$ -- dark matter particle mass}; (4)\rev{: $\Delta m_*$ -- old stellar particle mass}; (5)\rev{: $\rho_{\rm SF}/m_\mathrm{H}$ -- star formation density threshold normalised to the hydrogen mass}.
\\\vspace{0.1in}
}
\label{tab:resolution_table}
\end{table}

\begin{table*}
\centering
\caption{Summary of simulations}
\begin{tabular}{c@{\quad}cccc@{\quad\quad}cccccc}
\hline\hline
\\[-2ex]
Simulation name \rev{(1)} & \multicolumn{4}{c}{Initial conditions} & \multicolumn{6}{c}{Outcomes} \\ \hline
& Type \rev{(2)} & $B_0$ \rev{(3)} & $\Brms$ \rev{(4)} & Resolution \rev{(5)}  & $\langle B_\phi\rangle$ \rev{(6)} & $\sigma_B$ \rev{(7)}  & $\sigma_\mathrm{RM}$ \rev{(8)} & $\mathcal{I}_\mathrm{RM}$ \rev{(9)} & $p^+_\mathrm{RM}$ \rev{(10)} & $p^-_\mathrm{RM}$ \rev{(11)} \\
& & ($\mu$G) & ($\mu$G) & & ($\mu$G) & ($\mu$G) & (rad m$^{-2}$) & (rad m$^{-2}$)
\\[0.1ex]
\hline
\\[-2ex]
\texttt{I-0-2-L} & Isolated & 0 & 2 & Low & $\phantom{-}0.04$ & 0.34 & $25.47 \pm 1.01$ & $26.29 \pm 1.3$ & $0.52 \pm 0.03$ & $0.52 \pm 0.02$ \\
\texttt{I-2-0-L} & Isolated & 2 & 0 & Low & $\phantom{-}3.16$ & 1.73 & $66.12 \pm 1.88$ & $82.09 \pm 3.93$ & $0.86 \pm 0.02$ & $0.85 \pm 0.02$ \\
\texttt{I-2-2-L} & Isolated & 2 & 2 & Low & $\phantom{-}2.99$ & 1.59 &  $49.41 \pm 1.53$ & $57.2 \pm 3.04$ & $0.78 \pm 0.02$ & $0.77 \pm 0.02$ \\
\texttt{I-2-6-L} & Isolated & 2 & 6 & Low & $\phantom{-}2.83$ & 1.57 & $48.01 \pm 1.66$ & $52.6 \pm 2.82$ & $0.75 \pm 0.02$ & $0.7 \pm 0.02$ \\
\texttt{W-2-0-L} & Wind & 2 & 0 & Low & $\phantom{-}4.89$ & 1.79 & $100.84 \pm 3.1$ & $139.43 \pm 6.35$ & $0.9 \pm 0.02$ & $0.88 \pm 0.02$ \\
\texttt{W-2-2-L} & Wind & 2 & 2 & Low & $\phantom{-}3.21$ & 1.77 & $62.46 \pm 1.89$ & $74.43 \pm 3.35$ & $0.84 \pm 0.02$ & $0.79 \pm 0.02$ \\
\texttt{W-2-6-L} & Wind & 2 & 6 & Low & $\phantom{-}3.26$ & 1.78 & $55.2 \pm 1.55$ & $69.45 \pm 2.98$ & $0.84 \pm 0.02$ & $0.78 \pm 0.02$ \\
\texttt{I-0-12-M} & Isolated & 0 & 12 & Medium & $-0.10$ & $0.66$ & $27.55 \pm 1.0$ & $29.79 \pm 1.33$ & $0.52 \pm 0.02$ & $0.52 \pm 0.03$ \\
\texttt{I-2-6-M} & Isolated & 2 & 6 & Medium & 
$\phantom{-}1.20$ & $1.73$ & $33.17 \pm 1.25$ & $36.84 \pm 1.64$ & $0.68 \pm 0.02$ & $0.65 \pm 0.03$ \\
\texttt{W-2-6-M} & Wind & 2 & 6 & Medium & 
$\phantom{-}1.98$ & 1.94 & $40.48 \pm 1.33$ & $47.32 \pm 2.23$ & $0.71 \pm 0.02$ & $0.67 \pm 0.03$ \\
\texttt{I-2-6-H} & Isolated & 2 & 6 & High & 
$\phantom{-}0.99$ & 1.61 & $31.58 \pm 1.09$ & $35.0 \pm 1.56$ & $0.65 \pm 0.02$ & $0.6 \pm 0.03$
\\ \hline\hline
\end{tabular}\\[1.5ex]
\footnotesize{Columns: (1): \rev{Simulation name -- }name of simulation; (2): \rev{Initial conditions (Type) --} isolated or wind simulation; (3): \rev{Initial conditions ($B_0$) --} initial smooth azimuthal field strength; (4): \rev{Initial conditions ($\Brms$) --} initial turbulent field strength; (5): \rev{Initial conditions (Resolution) --} resolution (see \autoref{tab:resolution_table}); (6)\rev{: Outcomes ($\langle B_\phi\rangle$) --} final mean azimuthal field at two stellar scale lengths; (7): \rev{Outcomes ($\sigma_B$) --} final turbulent field strength at two stellar scale lengths; (8): \rev{Outcomes ($\sigma_\mathrm{RM}$) --} dispersion of RM; (9): \rev{Outcomes ($\mathcal{I}_\mathrm{RM}$) --} Interquartile (25$^\mathrm{th}$-75$^\mathrm{th}$ percentile) range; (10) and (11): \rev{$p^+_\mathrm{RM}$ and $p^-_\mathrm{RM}$ --} RM alignment fraction on the positive and negative RM sides of the galaxy\rev{, respectively}.
\\\vspace{0.1in}
}
\label{tab:master_table}
\end{table*}

\subsubsection{Isolated galaxy simulations}
\label{sssec:ioslated}

For our simulations of an isolated LMC-analogue, we use open boundary conditions and initialise a galactic disc with its centre of mass at the origin and zero net momentum. We initialise the dark matter and old stellar disc particles for simulations in this class using the \textsc{GALIC} code \citep{Yurin2014}. This code uses elements of Schwarzchild's technique, whereby one starts with some initial assignment of particle velocities and evolves it to optimise a merit function characterised by the difference between the target density field and the actual density response created by the N-body realisation associated with the initial velocities. Rather than optimising a global merit function as in a traditional Schwarzchild's method, this algorithm optimises the velocities of each particle directly, avoiding the need for an explicit orbit library. We set the parameters for \textsc{GALIC} to the same values adopted by \citet{Lucchini2020} for their simulations of the LMC. These are a DM halo with mass $M_{\rm 200} = 1.775 \times 10^{11} \Msun$ and a concentration parameter $c = 10$, and a stellar disc with mass $2.5 \times 10^9 \Msun$ and scale length and height $1.8 \kpc$ and $0.3 \kpc$, respectively. We use these parameters in \textsc{GALIC}'s model D1 \citep[see Table 1 in][]{Yurin2014}, which assumes a spherical dark halo, thin stellar disc, and axisymmetric velocity structure.

To initialise the gas disc, we set the total gas mass to $2.2 \times 10^9 \Msun$ and the disc scale length ($R_0$) and height ($Z_0$) are $4.8 \kpc$ and $0.48 \kpc$, respectively; we again take these parameters from \citet{Lucchini2020}. The gas density distribution is doubly-exponential, with the form 
\begin{align}\label{eq:rhogas}
    \rho = \rho_0 \exp\left(-\frac{r}{R_0}\right) \exp\left(-\frac{|z|}{Z_0}\right),
\end{align}
where $(r,\phi,z)$ are the standard cylindrical coordinates. In practice we implement this density distribution by drawing the initial positions of gas particles from a probability distribution function $p(r,\phi,z) \propto \rho$. Once we have drawn positions for all gas particles, we initialise their velocities by first computing the gravitational potential $\psi$ of the combined gas, stars, and dark matter using the \textsc{pytreegrav} package\footnote{\url{https://github.com/mikegrudic/pytreegrav}}\rev{. We} then \rev{set} particle velocities in the azimuthal direction $v_\phi$ such that all gas particles are in centrifugal equilibrium against gravity, i.e.,
\begin{align}
    \frac{v_\phi^2}{r} = -\frac{\partial \psi}{\partial r}.
\end{align}

We initialise the gas magnetic field with a combination of ordered and turbulent components. The ordered component is purely azimuthal, and follows the same functional form and scale length and height as the gas density:
\begin{align}\label{eq:rhoB}
    B_{\phi} = B_0 \exp\left(-\frac{r}{R_0}\right) \exp\left(-\frac{|z|}{Z_0}\right).
\end{align}
Here $B_0$ is the field strength, which varies from simulation to simulation as indicated in \autoref{tab:master_table}. While this field is analytically divergence-free, its discretisation onto particles introduces numerical divergence\rev{. We} remove this by carrying out an initial divergence cleaning and local approximate Hodge projection step prior to starting the simulation \citep{Hopkins2016divb}.  We generate the turbulent component of the field as a divergence-free Gaussian random field with a Kolmogorov spectrum of slope $-5/3$ \citep{Minter1996} and a standard deviation $\Brms$, which like $B_0$ varies between simulations as indicated in \autoref{tab:master_table}. The largest and smallest scales in our field are $1 \kpc$ and the resolution limit, respectively. We enforce zero divergence on this field using the projection method, whereby we set the radial component in Fourier space to zero.

\subsubsection{Simulations with ram pressure stripping}

We make several modifications to our initialisation procedure to simulate the LMC interacting with the MW CGM in our \texttt{W} simulations. First, we use a periodic domain of size $200\,\mathrm{kpc}\times 200\,\mathrm{kpc}\times\, 400\,\mathrm{kpc}$ rather than open boundary conditions as in the isolated simulations; the purpose of this change is to contain a background gas representing the MW CGM through which our LMC can move.

\begin{figure}
    \includegraphics[width=1\columnwidth]{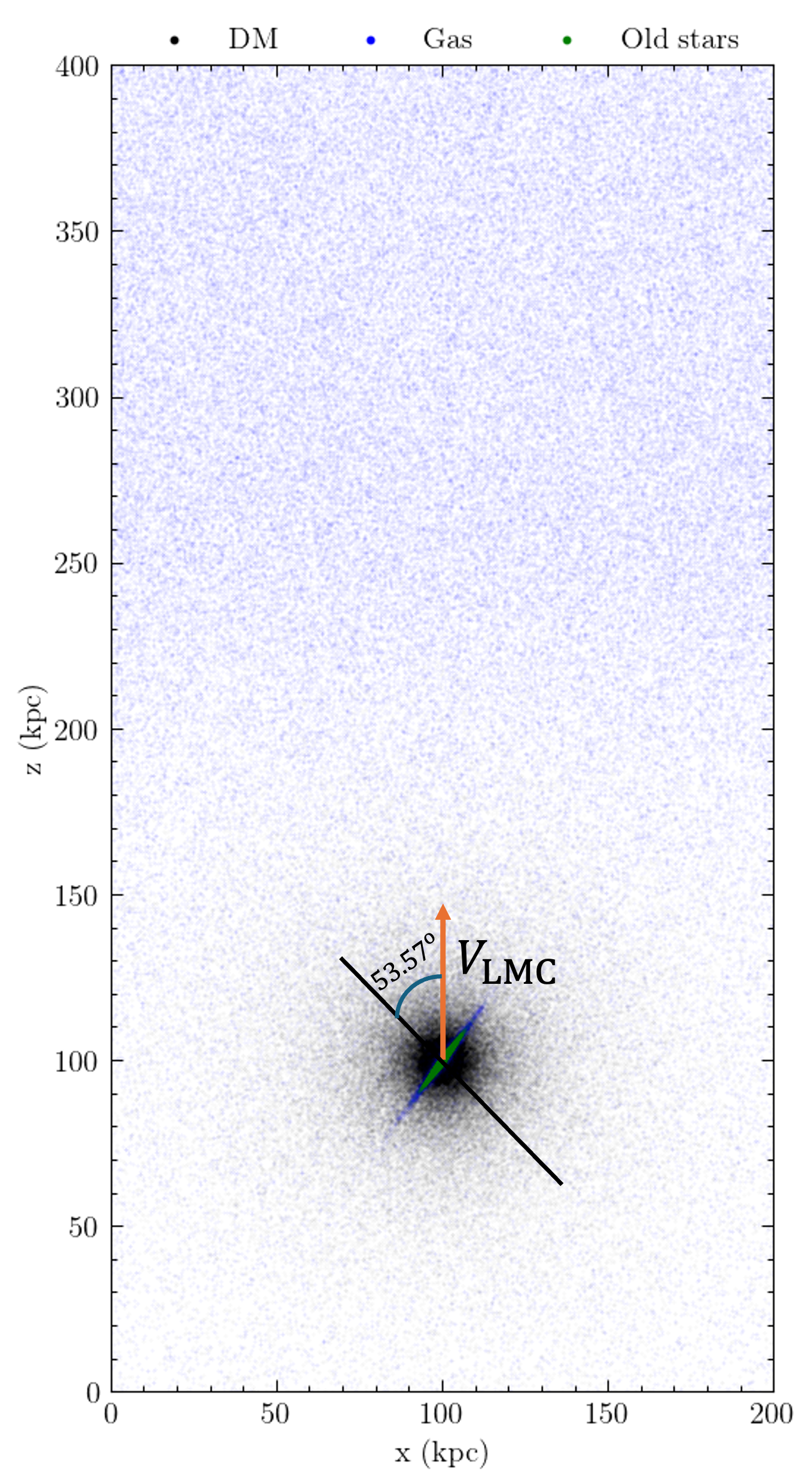} 
    \caption{Edge-on projection of initial particle positions in our \texttt{W} simulations. We show gas (blue), DM (black), and old stellar (green) particles; note that gas is present both in the LMC disc and in the background, and that the background has a density gradient chosen such that, as the simulated LMC moves, the density of material it encounters will reproduce the history of the real LMC. The angle of $53.57^{\circ}$ between the LMC velocity vector ($V_{\rm LMC} \hat{z}$) and angular momentum vector (direction illustrated by the black line) ensures that our simulated LMC experiences a wind with the same angle of attack as the real LMC.}
    \vspace{0.1in}
    \label{fig:wind_IC_annot}
\end{figure}

Our second major change is to the LMC's initial position and orientation, and is illustrated in \autoref{fig:wind_IC_annot}. We first initialise the LMC in exactly as in the \texttt{I} simulations; at this point the disc of the LMC lies in the $xy$-plane, and the angular momentum vector of the LMC is in the $z$ direction. We then translate the positions of all LMC gas, old stellar, and dark matter particles so that the centre of the LMC is at position $(x,y,z) = (100,100,100)$ kpc in our periodic domain, where the lower left corner of the domain is at the origin. We next rotate the particle positions and velocities by an angle of $\theta = 53.57^\circ$ counterclockwise in the $xz$ plane around the LMC centre, so that the angular momentum vector now points in a direction $\hat{n} = (-0.805, 0, 0.593)$, and add to each particle a velocity $V_\mathrm{LMC} \hat{z}$, where $V_\mathrm{LMC} = 320.7$ km s$^{-1}$. The values of $V_\mathrm{LMC}$ and $\theta$ are, respectively, the velocity of the LMC relative to the MW's centre and the angle between the LMC's velocity and angular momentum vectors \citep[their Table 3]{Salem2015}. Note that, over the 500 Myr of our simulation, the LMC will move 160 kpc, from $(100,100,100)$ kpc to $(100,100,260)$ kpc, and thus will remain far from the walls of our periodic domain.

Finally, we fill the simulation domain with a background gas whose density as a function of $z$-position $\rho(z)$ is chosen such that, as the LMC moves, the gas it encounters will have the same density as a function of time as the real LMC has encountered. We take this density as a function of time from the MW halo model of \citet{Salem2015}, who use a combination of models of the LMC's orbit and comparisons between \Hi~21 cm observations and simulations of ram pressure-stripped galaxies in order to estimate the density of the MW halo gas through which the LMC has passed over the last $\sim$Gyr. They find that this density reached a maximum $\nH = 1.1\times 10^{-4}$ cm$^{-3}$ during the LMC's pericentric approach $\sim 50$ Myr ago, but was significantly lower in the past when the LMC was further from the MW, and that as a result ram pressure stripping only became dynamically significant in the last $\sim 100-200$ Myr, significantly shorter than our simulation duration. To capture this change in density over time, we extract the background density as a function of time experienced by the LMC, $\rho_\mathrm{LMC}(t)$, from \citeauthor{Salem2015}'s Figure 4. We then set the density of the background gas in our simulation domain as $\rho(z) = \rho_\mathrm{LMC}(t(z))$, where
\begin{equation}
    t(z) = 
    \begin{cases}
        (z-z_0)/V_\mathrm{LMC} - t_0, & -z_0 < z - z_0 < V_\mathrm{LMC} t_0 \\
        0 & z - z_0 > V_\mathrm{LMC} t_0
    \end{cases},
\end{equation}
and $z_0 = 100$ kpc and $t_0 = 500$ Myr are the initial LMC $z$-position and run duration, respectively. The density variation $\rho(z)$ is visible in \autoref{fig:wind_IC_annot}. The background gas has zero initial velocity, and in order to ensure that it remains static prior to the passage of the LMC through it, we set the background temperature $T(z)$ such that $\rho(z)T(z)$ is constant and there is no pressure gradient; we normalise this relation such that $T = 5\times 10^6$ K at $z < z_0$, and given the density profile we take from \citeauthor{Salem2015} this implies a temperature range from $\approx 5\times 10^5 - 5\times 10^6$ K, reasonable CGM gas temperatures. However, \citeauthor{Salem2015}~also find that the CGM temperature has little effect on the process of ram pressure stripping. Finally, we initialise the magnetic field in the background gas using a Gaussian-random field with the same spectral shape as the turbulent disc field, but with a standard deviation of $0.6 \muG$ \citep{Pakmor2020, Shah2021, Heesen2023} and an outer scale of $30 \kpc$ \citep{Pakmor2020}.

The final minor differences in initialisation from the \texttt{I} case are intended to maintain continuity across the disc-halo interface. For densities, we define the disc-halo interface as the location where the gas disc density is the same as the gas halo density. We remove all the gas particles from the halo (disc) inside (outside) this interface to make densities continuous. 

For magnetic fields, the disc-halo interface is defined to be the location where the magnitude of the decaying, ordered component of disc magnetic field (\autoref{eq:rhoB}) is equal to the RMS strength of the halo magnetic field ($0.6$ $\mu$G). We set the disc field to zero outside this boundary and the halo field to zero inside it; note that this procedure does not guarantee that the magnetic field remains divergence-free at the disc-halo interface, but the initial divergence cleaning step carried out at simulation startup (\autoref{sssec:ioslated}) removes any divergence that it introduces. 

\subsection{Creating mock observables}
\label{sec:MockObservables}

\subsubsection{Definitions of the quantities}
\label{sec:definitions}

The three main observable diagnostics to which we have access when studying the LMC's magnetic field structure are dispersion measures (DMs), Faraday rotation measures (RMs), and emission measures (EMs). These three quantities are defined as
\begin{align}\label{eq:DM}
    {\rm DM} & =\int n_\mathrm{e} \, ds
    \\
    {\rm EM} & =\int n_\mathrm{e}^2 \, ds
    \label{eq:EM}
\end{align}
where \rev{the} integrals are evaluated along a path from the emitting source to the observer \rev{and RM is defined by \autoref{eq:rm_integral}.}
Observations of DM come primarily from sightlines to pulsars in the LMC, which are corrected for the contribution to the DM arising from the ISM of the MW
\citep{McCulloch1983, Seward1984, McConnell1991, Marshall1998, Crawford2001, Manchester2006, Ridley2013, Johnston2021}, while RMs are most often measured using linearly polarised synchrotron emission from background \rev{sources like AGNs}.
Emission measures are observed using the H$\alpha$ brightness. Note that, while the EM and DM are not directly sensitive to the magnetic field, they are useful for interpreting RMs, which mix together magnetic field and electron density; a common practice is to use DMs to infer typical electron densities, which are then in turn combined with RMs to yield magnetic field strengths.

\subsubsection{Geometry}

In order to generate mock observations, we must evaluate the integrals appearing in \autoref{eq:DM} - \autoref{eq:EM} through our simulation. To facilitate this exercise, we first transform our simulated LMC into a frame such that our \rev{LOS} is aligned with the $z$-axis. For \texttt{W} simulations this requires two steps: we first align the angular momentum vector of the LMC with the $z$ axis of the simulation by applying the rotation matrix 
\begin{align}
R_{\rm z} = 
\begin{bmatrix}
-0.42 & 0.71 & -0.57 \\
-0.42 & -0.71 & -0.57 \\
-0.8 & 0 & 0.6
\end{bmatrix}
\label{eq:Rz}
\end{align}
to the position of every gas particle, using positions measured relative to the centre of the simulated LMC. This rotation corresponds to $\alpha = 0^\circ$, $\beta = 53.57^\circ$, and $\gamma=225^\circ$, where $\alpha, \beta, \gamma$ are the Euler angles assuming extrinsic rotation. While $\gamma=225^\circ$ is not necessary to align the angular momentum vector of the LMC with $z$-axis, it is necessary to align the wind direction in the observer's frame when the second rotation is performed. We then rotate again to place the LMC into a frame where the observer is located at $z = +\infty$ (or outside the simulation domain) using the rotation matrix
\begin{align}
R_{\rm obs} = 
\begin{bmatrix}
0.64 & 0.63 & -0.44 \\
-0.76 & -0.53 & 0.37 \\
0 & -0.57 & -0.82
\end{bmatrix}.
\end{align}
The procedure for the \texttt{I} simulations simply uses the second step. We validate these transformations by comparing the wind velocity and angular momentum vectors with those documented by \citet{Salem2015} in their observer's frame.

\subsubsection{Dispersion measures}

Now that we have our observer's LOS aligned to the $\hat{z}$ direction, we can evaluate integrals through the simulation domain to generate mock DMs, RMs, and EMs. For DMs, since the observations to which we are comparing target pulsars within the LMC, we place in the simulations 200 synthetic pulsars, approximately equal to the steady-state number of \slug~clusters in the age range $5-10$ Myr expected for young pulsars. We choose the locations for these pulsars by starting from the positions of \slug~clusters (chosen at random if there are more clusters than pulsars) and adding a random displacement drawn from a 3d Gaussian distribution with a standard deviation of $500 \pc$. This displacement scale emerges from typical pulsar characteristics: spin-down times of 5-10 Myr \citep{Alvarez2004} and kick velocities of approximately 100 km s$^{-1}$ for the slower group of pulsars \citep{Verbunt2017}. Once we have our set of mock pulsar positions, we calculate DMs to them by evaluating the integral in \autoref{eq:DM} from the location of pulsar towards the observer along the \rev{LOS}.\footnote{Note that our \texttt{W} simulation domain includes some contribution from the medium we have placed in the box to represent the MW CGM, which in principle we should remove from the DM (and the RM) since the observations are corrected for the MW foreground. In practice, however, we find that the contribution of material far from the LMC is minimal because it represents only the local CGM environment experienced by the LMC during its orbit, rather than the complete MW CGM intervening between us and the LMC. We therefore omit this correction for simplicity.}

\subsubsection{Rotation and emission measures}

Since RMs are most often measured toward background quasars whose positions are random, we generate mock RM measurements by placing sightlines uniformly in a $10 \kpc \times 10 \kpc$ region around the LMC centre in the observer's frame and integrating \rev{\autoref{eq:rm_integral}} through the full domain. We pick 200 such sightlines as this is approximately equal to the number of RM sightlines available from the observations to which we will be comparing below, which come from  \citetalias{Gaensler2005}. We discuss the rationale for choosing this data set in more detail in \autoref{sec:RMObservations}. We find that choosing a different number of sightlines (100, 300, 400, 500) changes the outcome statistics by $\sim 10 \%$.

We also calculate EMs along the same lines of sight (\autoref{eq:EM}), which is important so that we can replicate the selection procedure used by \citetalias{Gaensler2005}. These authors discard any RM measurements along lines of sight with $\mathrm{EM} > 100$ pc / cm$^6$ to avoid observational errors due to depolarisation along lines of sight that pass through dense ionised regions. So that we can compare fairly with the observations, we replicate this cut by removing from our sample any RM measured along a line of with for which the EM is above this threshold.

Finally, in order to allow unbiased comparison with observations we also add realistic observational errors to our mock observables, and derive statistics from the real and mock observations using the same bootstrapping method to marginalise over those errors. We describe our procedure for doing so in \aref{ap:boot}.

\subsection{Estimating electron densities in the LMC}
\label{sec:PostProcess}

Our mock observables depend on the electron number density $n_\mathrm{e}$, and therefore generating realistic predictions requires careful consideration of the gas ionisation state, at a level beyond that provided by the ``quick and dirty'' model used in the on-the-fly simulation chemistry network. This first-order determination of the abundances of the various H and He ionisation states relies on the \grackle~tables, which assume a uniform UV background characteristic of intergalactic space \citep{HM2012}. This simple treatment overlooks the role of the Reynolds layer -- a dense ISM region that shields galaxies from background radiation emitted by distant AGN that pervades extragalactic space \citep{Reynolds1993}. Because of the presence of the Reynolds layer, the interstellar radiation field that prevails within galaxies is significantly softer than the field experienced by gas in the CGM. Our simulation includes both regions that are shielded by the Reynolds layer and regions outside it, and while the resulting spatial variation of the radiation field matters relatively little for the gas \textit{temperature} -- which is what is important for hydrodynamics, and thus why it is reasonable to use the simplifying assumption of a single radiation field when running the simulation -- it matters a great deal for the ionisation fractions of predominantly neutral gas, and thus in turn for the RM and DM signatures that this gas produces. Quantitatively, we find that for gas at the densities $\sim 0.1-1$ cm$^{-3}$ typical of the warm neutral medium, switching between a \citeauthor{HM2012} radiation field and one more appropriate for ISM conditions only produces $\sim 10\%$ differences in the equilibrium temperature, but can yield almost order of magnitude differences in the ionisation fraction, enough to strongly alter synthetic DM and RM signals.\footnote{The EM is much less sensitive to this effect, because its $n_e^2$ dependence means that the signal is dominated by gas that is almost completely ionised, and the choice of background radiation field only leads to substantial differences in gas that is mostly neutral.} We must therefore post-process the simulation data to take into account these effects and obtain a more realistic model of the gas ionisation state.

Our procedure for doing so is as follows. We first generate a set of tables for equilibrium ionisation states as a function of gas density and temperature using \textsc{Cloudy} \citep{Cloudy23}, following exactly the same procedure as for the \textsc{Grackle} library \citep{Smith2016}, but replacing the \citet{HM2012} radiation field used by \textsc{Grackle} with a background radiation field with a spectral energy distribution (SED) following \citet{Black1987}, subject to extinction by a cold neutral slab with column density $10^{21} \cm^{-2}$, typical for the ISM \citep{Stil2002, Draine2011, Saha2018}. In this calculation we also set a cosmic ray ionisation rate of $4.6 \times 10^{-16} \s^{-1}$ \citep{Indrirolo2007}. This radiation and cosmic ray field are typical of ISM conditions, as opposed to the CGM and IGM conditions assumed in \textsc{Grackle}. The output of this calculation is a set of ionisation states as a function of density and temperature in exactly the same format as the \textsc{Grackle} tables, but for an ISM radiation field.

We then assume that ionisation conditions switch from CGM-like to ISM-like at some characteristic transition temperature $T_\mathrm{Re}$ that characterises the Reynolds layer\rev{. We} also \rev{experiment} with placing the transition at a characteristic density and column density, and \rev{find} that the results are nearly identical. We will calibrate the value of $T_{\rm Re}$ below, but for the moment we simply assume we have a known value. Given this known value we now assign the ionisation state of each \gizmo~gas particle to the equilibrium value specified in the CGM-like or ISM-like table based on its temperature. Our final step is then to add in the effects of photionisation feedback. As noted above, we implement this in the simulations using a Str\"omgren volume method, as described by \citet{Armillotta2019}, following an earlier implementation from  \citet{Hopkins2018photo}. To ensure that our post-processed electron fractions are consistent with this treatment of photoionisation, we implement a post-processing algorithm to apply a similar Str\"omgren volume calculation to our outputs. The steps in this algorithm are as follows: (1) sort all the \slug~clusters from largest to smallest ionising luminosity; (2) sort all predominantly neutral ($n_e / n_\mathrm{H} < 0.5$) gas particles from smallest to largest distance to the most luminous star, with ionising photon production rate $\dot{N}_{\rm ion,*}$; (3) calculate the rate at which ionising photons must be supplied to maintain the nearest gas particle in a fully ionised state, $\Delta \dot{N}_{\rm ion, g} = \alpha_\mathrm{B} n_e n_\mathrm{H^+} V$, where $\alpha_\mathrm{B}$ is the case B recombination coefficient evaluated at $T=10^4$ K, $n_e$ and $n_\mathrm{H^+}$ are the number densities of electrons and protons assuming the particle is fully ionised, and $V$ is the particle volume; (4) if $\Delta \dot{N}_{\rm ion, g} < \dot{N}_{\rm ion,*}$, then (4a) change the chemical composition of the particle to fully ionised and remove it from the neutral particle list, deduct the photons available for the ionisation from the stellar budget (i.e., set $\dot{N}_{\rm ion,*}-\Delta\dot{N}_{\rm ion,g}\mapsto \dot{N}_{\rm ion,*}$), and go back to step (4); otherwise (4b) remove the \slug~cluster from the list of clusters and proceed to the next most luminous one, until none are left. At the end of this procedure, we have a description of the ionisation state that self-consistently accounts for both the background and local massive stellar radiation fields.

\begin{figure}
    \includegraphics[width=1\columnwidth]{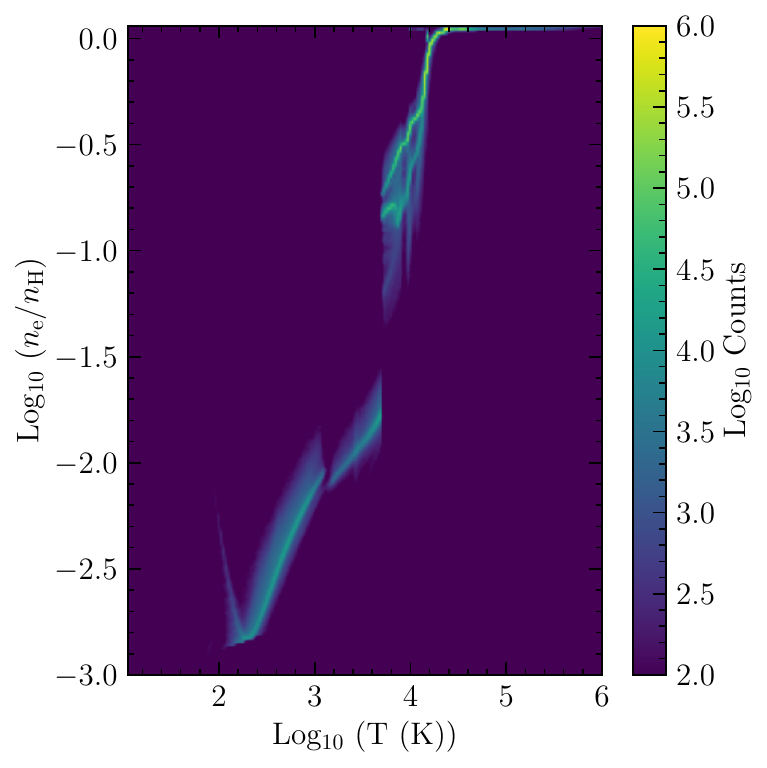} 
    \caption{Example 2D histogram of the distribution of gas particles in the plane defined by gas temperature and number of free electrons per H nucleon, computed using the procedure described in \autoref{sec:PostProcess}. This example is for the final time snapshot of simulation \texttt{W-2-6-M} computed using $T_\mathrm{Re} = 5000$ K.}
    \vspace{0.1in}
    \label{fig:phase_plot}
\end{figure}

\autoref{fig:phase_plot} shows an example of the joint distribution of free electrons per H nucleon, $n_\mathrm{e}/n_\mathrm{H}$, and temperature we compute via this procedure for the final snapshot of simulation \texttt{W-2-6-M}, using $T_\mathrm{Re} = 5000$ K. As expected, we see a clear break in the electron fraction at $T_\mathrm{Re}$ due to the change in radiation field at that temperature. Other features are essentially as expected: we see a band of high ionisation at $T = 10000-15000$ K driven by stellar photoionisation, and a transition to nearly complete ionisation above $\approx 20000$ K due to collisional ionisation. The small break at $\approx 3000$ K is due to the transition between the cold and warm neutral phases.

\begin{figure}
    \includegraphics[width=1\columnwidth]{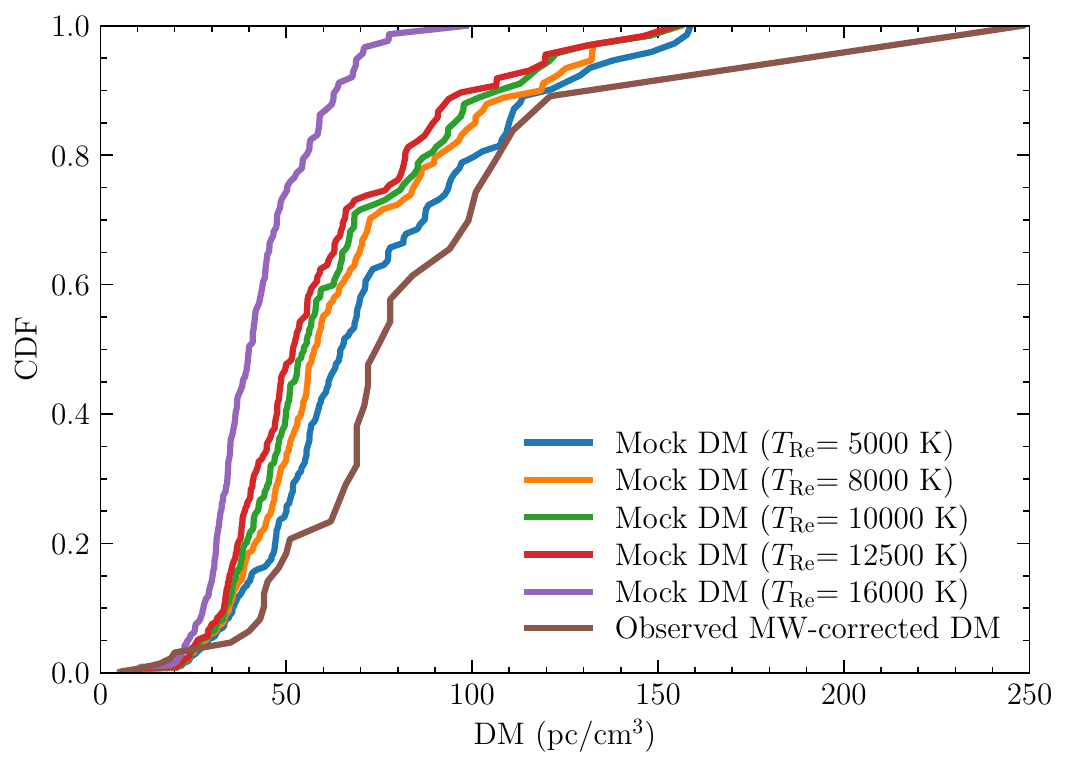} 
    \caption{CDF of observed MW-corrected DM (green) and the corresponding distribution of mock DMs generated from \texttt{W-2-6-M} at several different values of $T_\mathrm{Re}$, as indicated in the legend -- see text of \autoref{sec:PostProcess} for details. We see that, while the difference between the simulated CDFs is not strongly dependent of $T_\mathrm{Re}$, agreement is best for $T_{\rm Re} = 5000 \K$.}
    \vspace{0.1in}
    \label{fig:DM_LMC}
\end{figure}

We now seek to calibrate our choice of $T_\mathrm{Re}$ by matching our synthetic DMs to the distribution observed for LMC pulsars; we use DMs rather than RMs for this purpose because the DMs depend only on $n_\mathrm{e}$ and not on $\vec{B}$. We therefore repeat the calculation illustrated in \autoref{fig:phase_plot} for a grid of values $T_\mathrm{Re}/\mathrm{K} = (5000, 8000, 10000, 12500, 16000)$, again using simulation \texttt{W-2-6-M}; we choose this run because it is the most high-resolution case we have with wind. For each calculation, we generate synthetic DMs to the same set of synthetic pulsar positions following the procedure described in \autoref{sec:MockObservables}, and then plot the cumulative distribution function (CDF) or DMs, which we compare to the observed distribution taken from \citet{Johnston2021}. We show this comparison in \autoref{fig:DM_LMC}. As the plot shows, we obtain the best agreement for $T_\mathrm{Re} = 5000$ K, and we therefore adopt this value for the remainder of our analysis in this paper.

\section{Results}
\label{sec:results}
Here we report the results of our simulations. We first present a qualitative overview of the outcome to orient the reader in \autoref{sec:SimOverview}. We then carry out a detailed comparison of RM statistics between simulations and observations in \autoref{sec:RMObservations}. We explore the effects of ram pressure on gas properties and the 3D magnetic field structure, and the corresponding RM signatures, in \autoref{sec:3DRamLMC}. Finally, in \autoref{sec:GasPhase} we address the question of which gas phase is being probed by the RM observations.

\subsection{Overview of simulation outcomes}
\label{sec:SimOverview}

\begin{figure}
    \includegraphics[width=1\columnwidth]{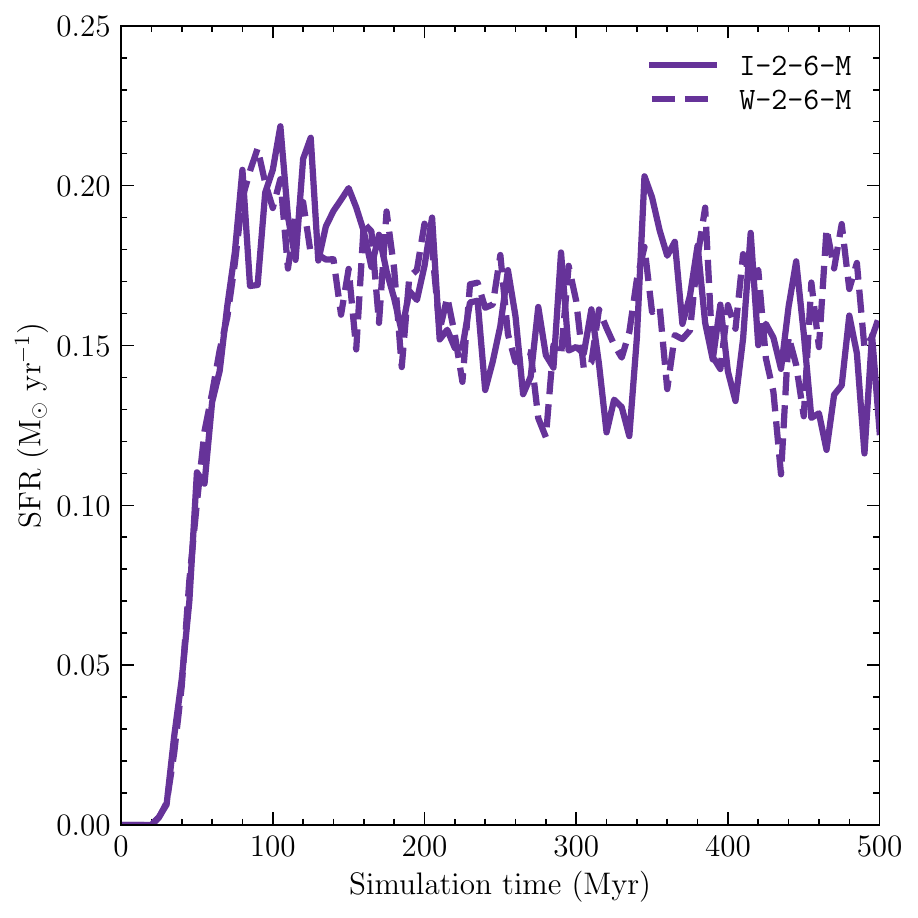} 
    \caption{Star formation rate (SFR) as a function of time in simulations \texttt{I-2-6-M} (solid) and \texttt{W-2-6-M} (dashed). We do not see any significant differences between the two cases, and the steady-state SFR of $0.1-0.2~$ M$_\odot$ yr$^{-1}$ is close to that of the observed LMC. \\}
    \label{fig:SFR_plot}
\end{figure}

\begin{figure*}
    \includegraphics[width=2\columnwidth]{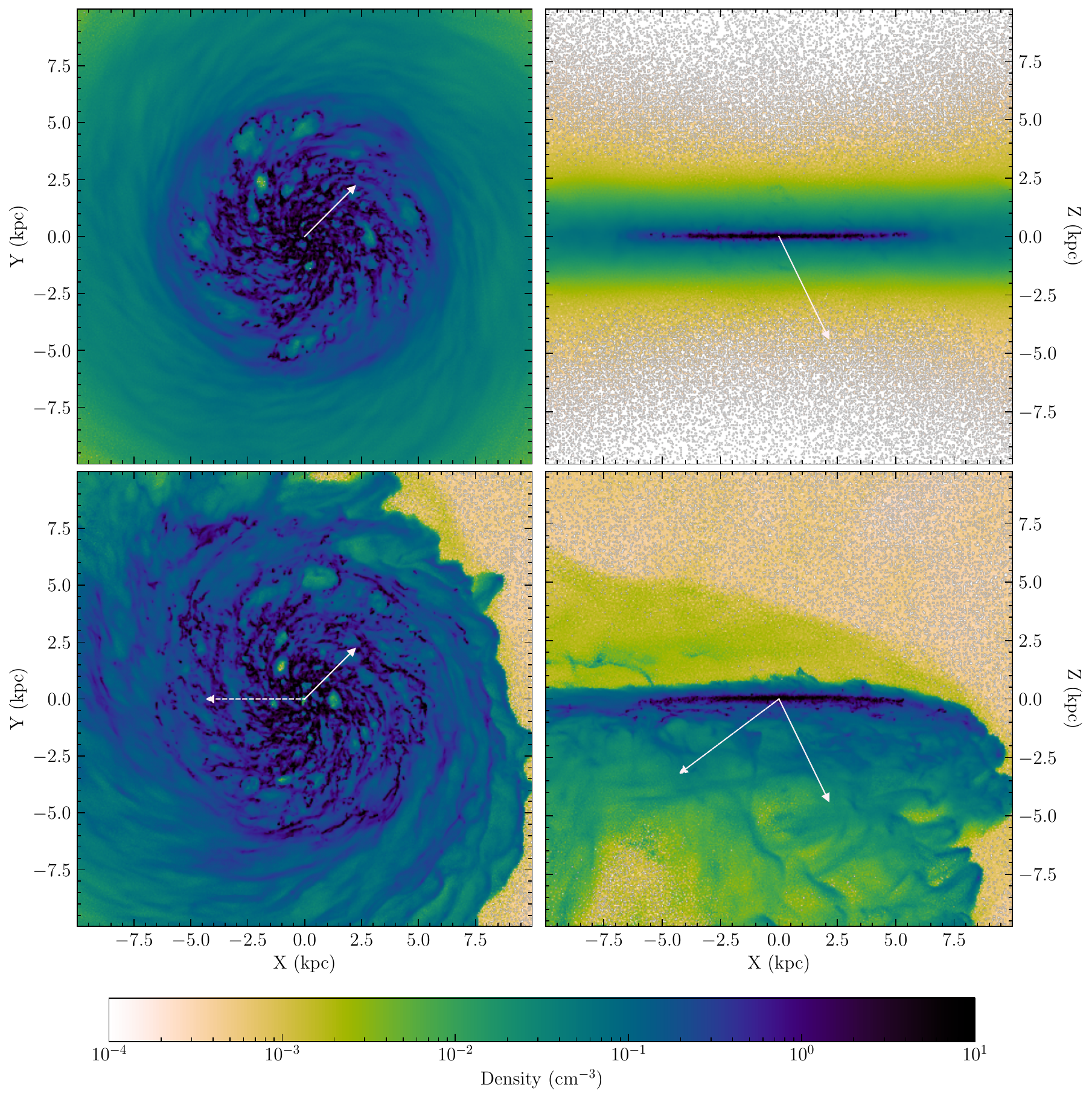} 
    \caption{Mass-weighted mean density along a LOS of $30 \kpc$ for simulations \texttt{I-2-6-M} (top) and \texttt{W-2-6-M} (bottom) after $500 \Myr$ of evolution. The left column shows a face-on projection, while the right column shows an edge-on projection; positions shown are relative to the LMC centre. We see that the presence of a wind in the \texttt{W} run adds dense turbulent eddies in the outskirts of the disc (face-on plots) and stripped material extending $\sim 15 \kpc$ off-plane (edge-on plots). These plots are in a frame where the LMC angular momentum vector is aligned with the $+z$ direction, and the wind direction is $-0.8 \hat{x}-0.6\hat{z}$, as indicated by the dashed white arrows. In the reference frame of the plot, the vector pointing from the LMC centre to the observer is $0.4\hat{x}-0.4\hat{y}-0.82\hat{z}$, and we show the projection of this vector onto the planes of the plot with the solid white arrow.\\}
    \label{fig:density_plot}
\end{figure*}

\begin{figure*}
    \includegraphics[width=2\columnwidth]{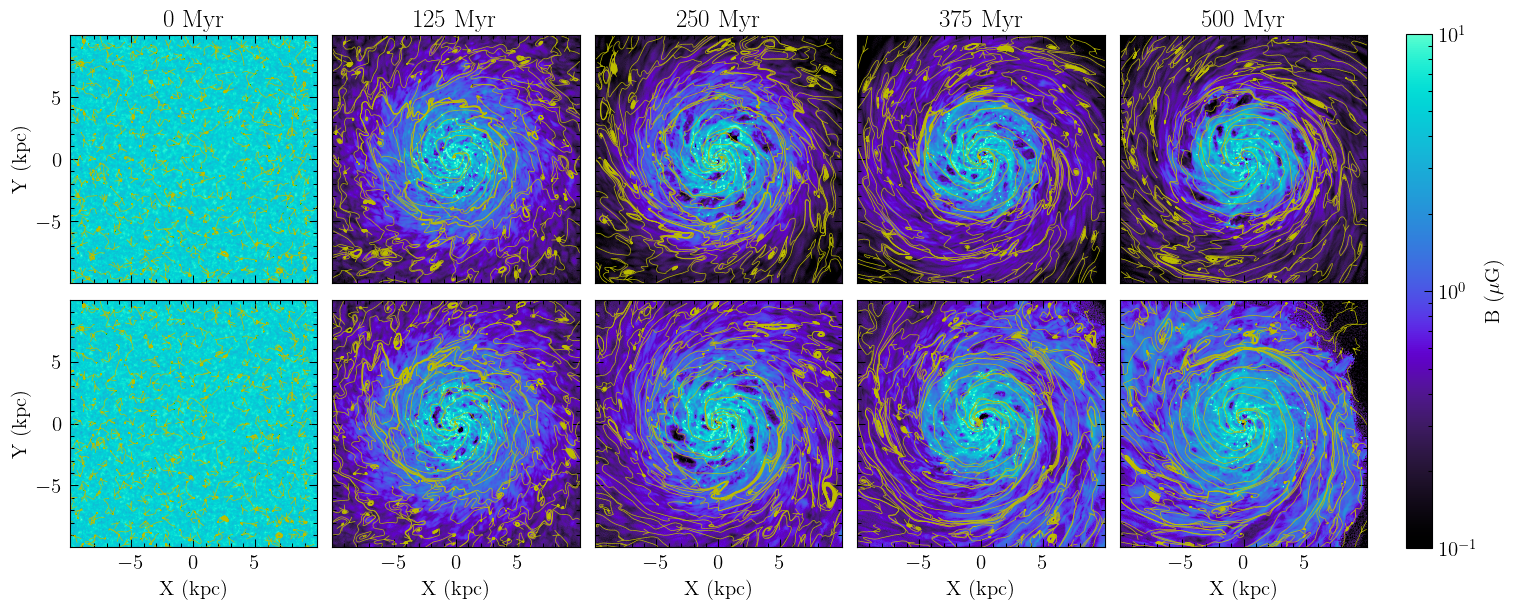} 
    \caption{Density-weighted mean magnetic field as a function of time in the simulation (columns from left to right) for example isolated \texttt{I-2-6-M} (top) and wind \texttt{W-2-6-M} (bottom) simulations. The yellow overlaid curves show the direction of magnetic field lines while the background colour shows field magnitude. We see that the simulation evolves from predominantly turbulent initial fields to relatively ordered final fields. At late times we observe major differences in both the morphology and the strength of fields between the isolated and the wind cases.\\}
    \label{fig:B_plot}
\end{figure*}

All our simulations undergo an initial burst of star formation before settling to a relatively steady-state regime after $\approx 150$ Myr. We show examples of the star formation history for two example simulations in \autoref{fig:SFR_plot}, but the results for all simulations are qualitatively similar, with no strong dependence on the initial magnetic field configuration or presence or absence of a wind from the MW CGM. In steady-state, the SFRs in our simulations are $\approx 0.1-0.2$ M$_\odot$ yr$^{-1}$, in good agreement with observational estimates for the LMC \citep[e.g.,][]{Maschberger2010, Jameson16a}.

\autoref{fig:density_plot} shows the mass-weighted mean density averaged in face-on and edge-on projection through the simulated LMC after 500 Myr of evolution, comparing two example simulations, one isolated (\texttt{I-2-6-M}) and with a wind (\texttt{W-2-6-M}); we select these two because their overall outcomes are reasonably typical of other isolated and wind cases. For the purpose of making this plot for the \texttt{W} simulations, we rotate particle position about the LMC centre by the rotation matrix 
\begin{equation}
R = 
\begin{bmatrix}
0.59 & 0 & 0.8 \\
0 & 1 & 0 \\
-0.8 & 0 & 59
\end{bmatrix},
\end{equation}
which is the inverse of the rotation by $53.57^\circ$ we apply to the initial conditions; this serves to rotate the LMC back to an orientation where its angular momentum vector lies in the $\hat{z}$ direction. In terms of Euler angles, this is equivalent to $\alpha = 0^\circ$, $\beta = 53.57^\circ$, and $\gamma=0^\circ$.\footnote{Note that this is slightly different from the rotation used in \autoref{eq:Rz}, which has $\gamma=255^\circ$; both rotations align the angular momentum vector with the $z$ axis, but the one given in \autoref{eq:Rz} also applies an additional rotation about the $z$ axis so that the wind velocity vector is properly aligned relative to the \rev{LOS} from the Milky Way to the LMC.}

We see from \autoref{fig:density_plot} that the dense interstellar medium of our simulated LMC extends approximately 5 kpc from the galactic centre, consistent with HI observations \citep{Kim1998, StaveleySmith2003}. Supernova feedback introduces turbulence that creates a flocculent structure throughout the ISM. The isolated case contains low-density material extending to larger radii, but in the \texttt{W} case ram pressure stripping has removed much of the diffuse material. Removal begins with the hot, diffuse gas that is more weakly gravitationally bound \citep{Balogh2000}, and then advances outside-in as the simulation progresses, consistent with observational inferences from truncated discs  \citep{Warmels1988, Cayatte1990, Vollmer2001, Lee2022}. The ram pressure, however, is insufficient to remove the denser ISM components, due to the LMC's substantial mass, relatively modest velocity through the MW CGM, and the relatively low ambient CGM density.  In the \texttt{W} case, the edge-on projection reveals a dense tail extending $\sim 15 \kpc$ below and to the left of the LMC (in the orientation of the projection shown), marking the trajectory of the stripped material and giving it a jellyfish-like appearance; as shown in the \rev{\autoref{fig:density_plot}}, our \rev{LOS} to the LMC intersects this tail of material, which means that it can contribute to the \rev{LOS} integrated EM, DM, and RM. There is a corresponding bow shock above and to the right  \citep{Bekki2009, Ebeling2014, McPartland2015, Rohr2023}. Ram pressure can compress the gas, especially on the leading edge, and we see marginal evidence for this towards the leading edge in the off-plane direction. 

\autoref{fig:B_plot} shows the time evolution of the magnetic fields for the same two runs as shown in \autoref{fig:density_plot} at intervals of $125 \Myr$. Since the initial turbulent field ($6 \muG$) is 3 times stronger than the initial smooth field ($2 \muG$), the leftmost panels showing the initial conditions contain mostly randomly oriented magnetic field vectors, as expected. However, the rotation of the galaxy quickly orders this into predominantly azimuthal fields. Field strengths in the galactic centre rise rapidly due to strong compression caused by cooling, so that in steady state the magnetic field and density structures are strongly correlated.

The magnetic field evolution in our isolated (\texttt{I-2-6-M}) and wind (\texttt{W-2-6-M}) runs remain similar until 125 Myr, after which significant differences emerge in the outer regions ($r \gtrsim 5$ kpc). At 500 Myr, while the isolated case shows weak magnetisation in the outskirts, the wind case exhibits stronger fields both in the outer regions and central 5 kpc. This enhancement stems from two effects: ram pressure compression leading to density and magnetic field amplification in central regions, and the LMC sweeping up outflows and magnetised circumgalactic medium (CGM) material in the outer regions. The effect intensifies over time as the LMC traverses increasingly dense CGM regions (see \autoref{fig:wind_IC_annot}), leading to disc truncation on the leading edge while maintaining overall stronger magnetic fields. The magnetic field magnitude shows a strong correlation with gas density, as can be seen by comparing the bottom left panel of \autoref{fig:density_plot} with the bottom right panel of \autoref{fig:B_plot}. Thus, a dense magnetised tail exists towards the trailing edge of the LMC. These features suggest that ram pressure stripping should leave observable signatures in RM maps of the LMC.

\subsection{Comparison with RM observations}
\label{sec:RMObservations}

Having given a general overview of the outcome of our simulations, we now proceed with our primary goal, which is a quantitative comparison between our mock RM data for all simulations and real observations. For the latter, there have been two major studies of interest that report large samples of RM measurements through the LMC to polarised extragalactic background sources:
\citetalias{Gaensler2005} and \citet{Livingston2024}. Both use 1.4 GHz data from the Australia Telescope Compact Array (ATCA). While the more recent measurements from \citeauthor{Livingston2024}~offer a larger number of sources and thus higher spatial resolution, for the purposes of a statistical comparison we prefer to use the \citetalias{Gaensler2005} data. The reason is that these data, which were collected as part of the hydrogen line survey \citep{Kim1998} and include 291 background sources across a 130 square degree field (of which roughly 100 cover the LMC), are randomly selected and thus offer a statistically-unbiased sampling of the LMC's magnetic field\footnote{There remain biases in \textit{detection} of polarisation, as opposed to target selection, but these are easier to model, and we will do so below.}. By contrast, \citeauthor{Livingston2024} specifically chose sightlines toward regions of high \Hi~column density, with the primary goal of detecting \Hi~absorption. Since this selection was done by hand, there is no easy way for us to reproduce it in mock observations, and we therefore focus our statistical comparison of simulations and observations on the uniformly-sampled \citetalias{Gaensler2005} map despite its lower resolution. We will make use of the foreground-subtracted RM measurements from this data set for the remainder of this section.

\begin{figure*}
    \includegraphics[width=2\columnwidth]{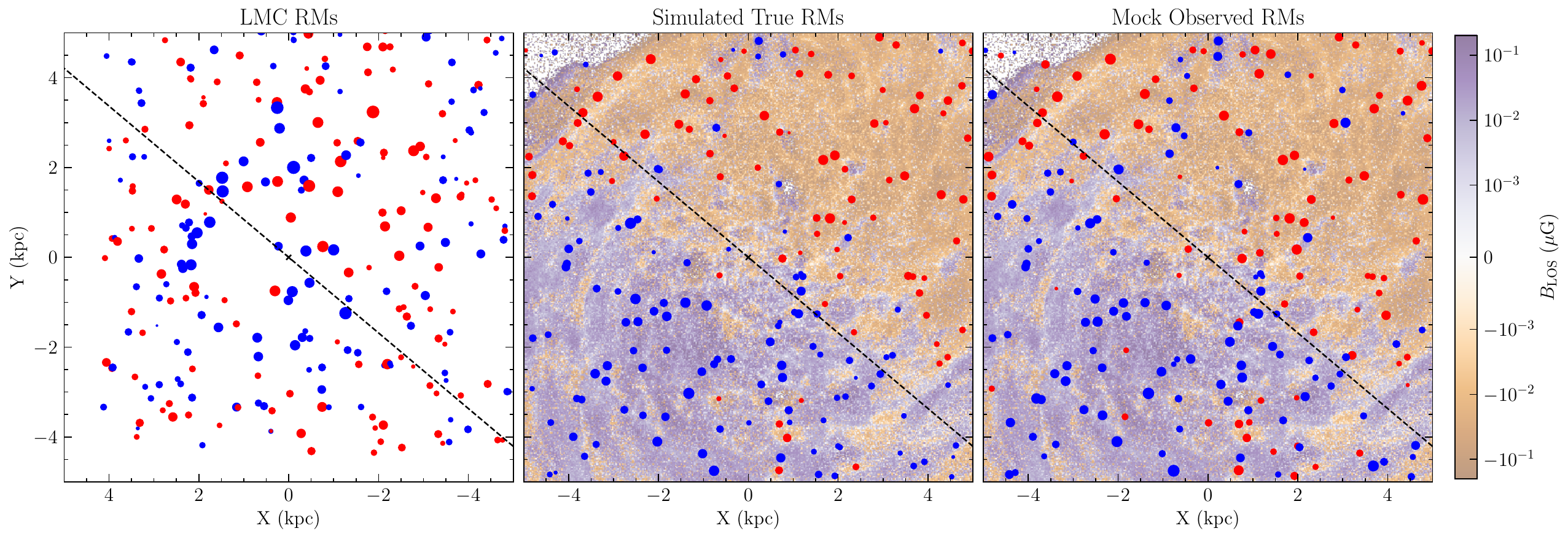} 
    \caption{The \citetalias{Gaensler2005} RM map (left), a sample mock RM with central RM values map from run \texttt{W-2-6-M} (middle), and a sample mock RM map in a realisation where we include simulated observational errors from run \texttt{W-2-6-M} (right). In the leftmost panel, blue (red) circles show positive (negative) RMs, with the size of the circle is $\propto \mathrm{|RM|}^{0.5}$, with the largest circle having a magnitude of $\sim 250$ rad/m$^2$. The dashed lines show the magnetic line of nodes, which is perpendicular to the line of nodes of the LMC's rotation as seen from the Sun, computed using the LMC geometry and kinematic model of \citet{vdm2004}. If the LMC's magnetic field were purely azimuthal, RMs should flip signs from positive to negative across this line. The mock RM maps in the rightmost two panels are overlaid on a background showing the true volume-weighted mean LOS ($10$ kpc depth) magnetic field. Comparing the middle and the right panel shows that the inclusion of simulated observational error in the treatment of mock RM data can change the value and alignment fractions of the dataset.\\}
    \label{fig:RM_plot}
\end{figure*}

\begin{figure}
    \includegraphics[width=1\columnwidth]{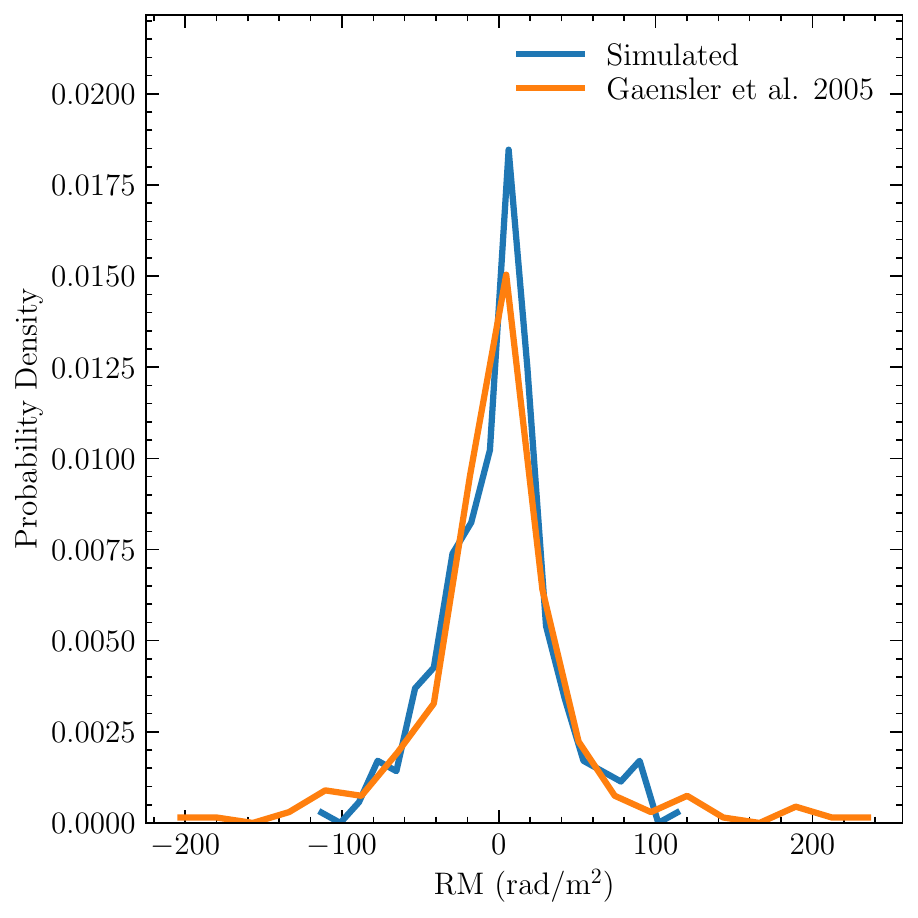} 
    \caption{PDFs of RMs from the two leftmost maps shown in \autoref{fig:RM_plot}. The comparison of observed and mock PDFs demonstrates that the simulation \texttt{W-2-6-M} does a good job of reproducing the observed LMC RM distribution.\\}
    \label{fig:RM_compare}
\end{figure}

The left panel of \autoref{fig:RM_plot} shows the \citetalias{Gaensler2005} RM map of the LMC, while the middle and right panels show mock RM maps from the final time snapshot of our simulations (computed as described in \autoref{sec:MockObservables}) for comparison. The middle panel shows the central RM values, and the right panel shows one of our realisations with our treatment of observational errors in the mock data (see \aref{ap:boot} for details). In comparing our simulations to these measurements, we focus on three statistical quantities. The first two are the interquartile (25th - 75th percentile) range and the dispersion of RM values, $\mathcal{I}_\mathrm{RM}$ and $\sigma_\mathrm{RM}$, which characterise the typical strength of the LMC's magnetic field (along with the electron density). The IQR is more robust against outliers and thus we will give slightly more weight to it in our analysis, but we will report both $\mathcal{I}_\mathrm{RM}$ and $\sigma_\mathrm{RM}$ in what follows.
Our third statistic is an alignment fraction metric, which we use to characterise the fraction of RMs that have the ``wrong'' sign. In \autoref{fig:RM_plot}, the dashed lines shown in each panel are drawn to be perpendicular to the LMC's line-of-nodes as determined from the rotation parameters of \citet{vdm2004}. If the LMC's magnetic field were purely azimuthal, the direction of the \rev{LOS} magnetic field would change sign across this line; if the magnetic field direction were parallel to the LMC's direction of rotation, given its orientation on the plane of the sky, we would expect regions to the left of the dashed line to show purely positive RM values, while regions to the right would have negative values. In fact, as \autoref{fig:RM_plot} shows, while there is a trend in this direction, it is clearly noisy, with significant numbers of RM measurements showing a sign opposite to the one we would expect for a simple azimuthal field geometry; the extent of this misalignment therefore characterises the extent to which the magnetic field displays turbulent structure that deviates from a simple ordered, azimuthal geometry. We quantify this by defining the alignment fractions $p^+_\mathrm{RM}$ and $p^-_\mathrm{RM}$ as the fraction of positive and negative RMs on their expected sides, i.e., $p^+_\mathrm{RM}$ represents the fraction of measurements to the right of the line of nodes shown in the left panel of \autoref{fig:RM_plot} that do in fact show positive RM, and similarly for $p^-_\mathrm{RM}$. A completely random RM distribution would have 50\% alignment fraction, while a purely azimuthal field would have 100\%.



We compute our three statistics from the mock and real observations using the analysis method outlined in \aref{ap:boot} to properly marginalise over the uncertainties. The effect of these uncertainties is clear from the middle and right panels of \autoref{fig:RM_plot}, where we see different RM values with more misalignment in the realisation that accounts for the observational errors. For the observations, this analysis yields $\sigma_\mathrm{RM} = 55.54 \pm 1.98$ rad/m$^2$, $\mathcal{I}_\mathrm{RM}=48.5 \pm 2.13$ rad/m$^2$, $p_\mathrm{RM}^+ = 0.52 \pm 0.02$, and $p_\mathrm{RM}^- = 0.54 \pm 0.02$; for the mock observations, we analyse the final time snapshot of each run, and report the resulting values in  \autoref{tab:resolution_table}. Here we can start to see the effects of different initial magnetic fields on the observables like the RM: when we initialise the simulation with purely turbulent magnetic fields (as is the case for runs \texttt{I-0-2-L} and \texttt{I-0-12-M}), the alignment fractions are close to $0.5$, meaning a randomly distributed RM map. However, the typical magnetic field strengths we find at the ends of these runs are too small due to efficient cancellation of RM contributions from regions in which the field is oppositely-directed, leading to $\sigma_{\rm RM}$ and $\mathcal{I}_\mathrm{RM}$ values that are too small compared to what we observe. When we initialise the magnetic fields to be purely azimuthal (\texttt{I-2-0-L}, \texttt{W-2-0-L}), the alignment fractions ($\sim 0.85, 0.89$), $\sigma_{\rm RM}$ ($\sim 66, 101~\rad/\m^2$), and $\mathcal{I}_\mathrm{RM}$ ($\sim 82, 139~\rad/\m^2$) are too high compared to the observations; the high $\sigma_\mathrm{RM}$ and $\mathcal{I}_\mathrm{RM}$ values occur because the ordered field amplifies significantly as the simulation runs, yielding high RM magnitudes in both the positive and negative directions. This makes it clear that recreating the observed alignment fractions, $\sigma_{\rm RM}$, and $\mathcal{I}_\mathrm{RM}$ will require a mix of turbulent and ordered fields.

Examining the various mixed-field values we have tried, we find that the case with a wind where the turbulent fields are 3 times stronger than the ordered fields in the initial conditions -- run \texttt{W-2-6-M} -- reproduces the dispersion of RMs, interquartile range of RMs, and the alignment fraction reasonably well. We show the distribution of RM values for this run in \autoref{fig:RM_compare}, demonstrating that this run provides a very good match to the observations. There is an excellent match between the observed $\mathcal{I}_\mathrm{RM} = 48.5 \pm 2.13$ rad/m$^2$ and \texttt{W-2-6-M} $\mathcal{I}_\mathrm{RM} = 47.32 \pm 2.23$ rad/m$^2$; agreement is slightly less good for $\sigma_{\rm RM}$ ($40.48 \pm 1.33$ rad/m$^2$ simulated versus $55.54\pm 1.98$ rad/m$^2$ observed), but since this quantity is not as robust as $\mathcal{I}_\mathrm{RM}$ against outliers, and we can be seen in \autoref{fig:RM_compare} that a number of outliers with very high RM are present in the data, we consider this run a good match overall to the observations. Most of the subsequent scientific discussion will therefore revolve around this case unless otherwise stated.

\rev{Before moving on, however, it is helpful to pause and compare our results to those of \citet{bustard2020}, whose simulations are the most similar to ours in the existing literature in that they include both magnetic fields and the wind from the Milky Way CGM. The simulations also differ in important respects: they include cosmic ray transport, which we do not, but we have much higher resolution, a more realistic treatment of ISM thermodynamics and ionisation, self-gravity, and self-consistent star formation and stellar feedback. They also initialise their simulations with a purely toroidal field with a peak strength of 4 $\mu$G in the disc, with no turbulent component and no magnetic fields in the CGM. Despite these differences, their RM maps look qualitatively similar to ours (compare their Figure 10 to our \autoref{fig:RM_plot}), although it seems unlikely that they would agree at the level of detailed statistics given our finding that a significant initial turbulent field is required to reproduce the data.}

\subsection{On the importance of ram pressure for LMC rotation measures}
\label{sec:3DRamLMC}

\begin{figure}
    \includegraphics[width=1\columnwidth]{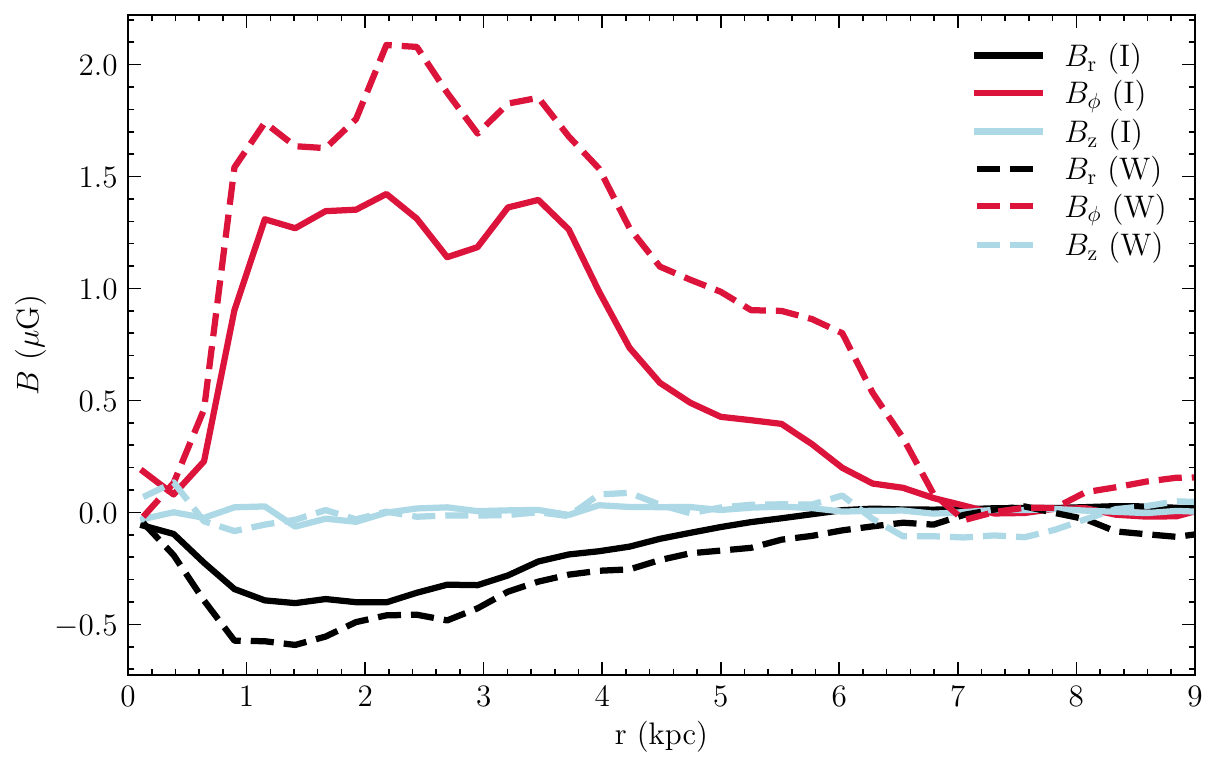} 
    \caption{Volume-weighted mean magnetic fields averaged over thin cylindrical shells with thickness $0.25$ kpc and height $0.5$ kpc in runs \texttt{I-2-6-M} (solid lines) and \texttt{W-2-6-M} (dashed lines), which are identical except that the latter includes the wind due to the MW halo and the former does not. Different colours show the three components of the field in a cylindrical coordinate system where the angular momentum axis of the simulated LMC is oriented in the $\hat{z}$ direction. There is an evident increase of almost $50\%$ in the azimuthal ($B_{\rm \phi}$) and radial ($B_{\rm r}$) magnetic field strengths in the presence of wind.\\}
    \label{fig:Brad_profile}
\end{figure}

\begin{figure*}
    \includegraphics[width=2\columnwidth]{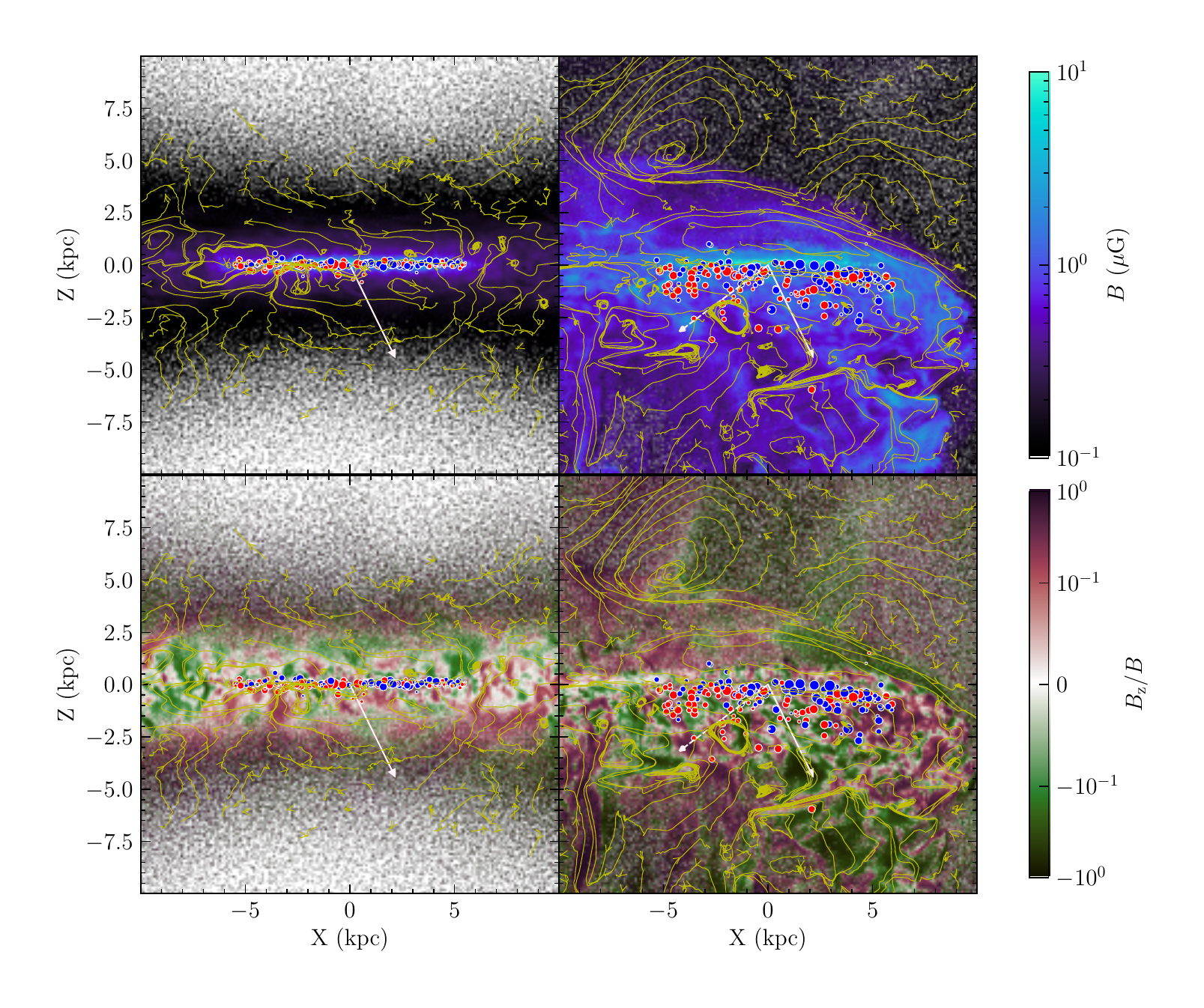} 
    \caption{The top two panels show the density-weighted mean magnetic field strength averaged over a $30$ kpc thick slab centred on the LMC centre for runs \texttt{I-2-6-M} (isolated case, left) and \texttt{W-2-6-M} (wind case, right) at the end of the simulation, in a frame where the LMC angular momentum vector is aligned with the $+z$ direction and the wind direction is $-0.8 \hat{x}-0.6\hat{z}$, as indicated by the dashed white arrow. The bottom two panels show the ratio of vertical to total magnetic fields over the same region in the two runs. Yellow lines, which are the same in both panels, show magnetic field lines in the plane of the plot. In the reference frame of the plot, the vector pointing from the LMC centre to the observer is $0.4\hat{x}-0.4\hat{y}-0.82\hat{z}$, and we show the projection of this vector onto the plane of the plot with 
    the solid white arrow. Blue (positive RM) and red (negative RM) circles mark the RM-weighted mean position along each of our 200 sample lines of sight (\autoref{eq:zLOS}) projected onto the plane of the plot. The sizes of the circles are scaled by $\rm |RM|^{0.5}$. We see that, in the run including a wind from the LMC's passage through the MW CGM (right panels), the wind drags out the RM signature from the plane of the LMC to $\sim 1$ kpc off-plane in the direction of the tailwind and the observer.\\}
    \label{fig:B_edgeplot}
\end{figure*}

Now that we have identified a simulation that reasonably reproduces observations of the LMC, we can use it to understand the physical origins of features of those observations. We first do so with respect to the effects of ram pressure from the MW halo on the LMC, which we \rev{again} assess by comparing our wind simulation that best matches the data (\texttt{W-2-6-M}) with the corresponding simulation with identical initial conditions in the LMC disc but without the MW halo (\texttt{I-2-6-M}), which matches the data less well.

\autoref{fig:Brad_profile} shows the volume-weighted mean magnetic fields in cylindrical shells of height $0.5$ kpc and thickness $0.25$ kpc as a function of cylindrical radius in these two cases. The vertical ($B_z$) component of the magnetic field is negligible in both cases, meaning that in both, the field largely lies in the plane of the disc. Interestingly, the radial and azimuthal magnetic field components are $\sim 50 \%$ larger in the run including the MW halo wind, despite the field strength starting at identical values in the two simulations. One might be tempted to attribute this to gas compression leading to magnetic field amplification in the wind run, but we find that the relative increase in density in the wind run is significantly smaller than the amplification in magnetic field strength, i.e., $\rho_{\texttt{W}}/\rho_{\texttt{I}}<B_{\texttt{W}}/B_{\texttt{I}}$ at all radii, particularly in the outskirts ($r>4$ kpc). Since even if the compression due to the wind is purely orthogonal to the field (the most favourable orientation) the field amplification is only linear in density, i.e., $B \propto \rho$, this implies that compression alone cannot explain the magnetic field enhancement we measure.

Instead a more likely explanation for the enhanced field in \texttt{W-2-6-M} compared to \texttt{I-2-6-M} is that when the LMC traverses the magnetised MW CGM, it sweeps up material which gets deposited in its ISM and increases the field strength there. This effect becomes particularly evident past $\sim 200 \Myr$ in our simulation. This time is special for the following reason: the LMC is at all times travelling supersonically with respect to gas in the halo and thus drives a bow shock, but prior to this time, the halo material through which the LMC passes is sufficiently low density that the LMC is supersonic but not super-Alfv\'enic. In this state, the passage of the LMC through the halo is mediated by fast magnetosonic waves that run ahead of the galaxy and create a standoff between the galaxy and the bow shock. As the LMC traverses denser and denser material over time, however, the halo Alfv\'en speed drops and the standoff distance decreases, until eventually, around 200 Myr the LMC catches up with the strongly magnetised bow shock and accretes the swept-up material and the magnetic field associated with it.

\autoref{fig:B_edgeplot} shows slices through the magnetic field structure of the LMC and illustrates this phenomenon. The left panels show run \texttt{I-2-6-M}, and demonstrate that, in the absence of a wind, the galaxy has a mostly symmetric magnetic field structure across the disc plane. The top left panel shows that the field strength falls rapidly away from the midplane, and the bottom left panel shows that, in the disc plane where the field is strongest, the field is oriented mostly horizontally, residing in the disc plane. Immediately above and below the plane ($|z|\geq 200$ pc), the field is turbulent, with no strong preference for a particular vertical component.

In contrast, the right panels of the figure show \texttt{W-2-6-M}, which has starkly different features, particularly away from the disc plane. First, there is a very strong asymmetry above and below the disc due to the wind being directed towards the bottom left of the figure (as shown by the white dashed arrow). As discussed above, this motion creates a bow shock above the disc with enhanced magnetic fields parallel to the interface between the background CGM and LMC material. The relative increase in the magnetic field magnitude and its alignment along the boundary is due to magnetic draping, a phenomenon where the magnetic fields are swept up and slip along the surface of a moving object such as a galaxy or a cloud as it interacts with the surrounding medium (e.g., see Figure 3 of \citealt{Pfrommer2010}, among many other examples -- \citealt{Lyutikov2006, Ruszkowski2007, Dursi2008, Ruszkowski2014, Sparre2020, Jung2023}). The region below the disc consists of a magnetised gas tail, extending $\sim 15 \kpc$ off-plane, created by the material stripped from the LMC ISM. As a result of these effects, we see that the magnetic field in \texttt{W-2-6-M} is both stronger and more organised than in \texttt{I-2-6-M}.

These changes in the magnetic configuration have dramatic effects on the observed rotation measures. To quantify these effects, we begin by defining the RM-weighted mean position along any given \rev{LOS} $i$ through the LMC as
\begin{equation}
        \langle s \rangle_{\rm RM, i} = \frac{\int s n_\mathrm{e} |\vec{B}\cdot \hat{s}| \, ds}
        {\int n_\mathrm{e} |\vec{B}\cdot \hat{s}| \, ds},\label{eq:zLOS}
\end{equation}
where $s$ is a coordinate measuring the position along the \rev{LOS} (c.f.~\autoref{sec:definitions}), and we place $s=0$ at the position where the LOS passes through the LMC plane and $s=\infty$ at the location of the observer. The factor $n_\mathrm{e} |\vec{B}\cdot\hat{s}|$ appearing in this expression is simply the absolute value of the differential contribution made by the material at any given point $s$ to the RM. \rev{Therefore,} $\langle s\rangle_{\mathrm{RM},i}$ can be used to characterise the average position along this \rev{LOS} from which the RM signal comes.


\begin{figure}
    \includegraphics[width=1\columnwidth]{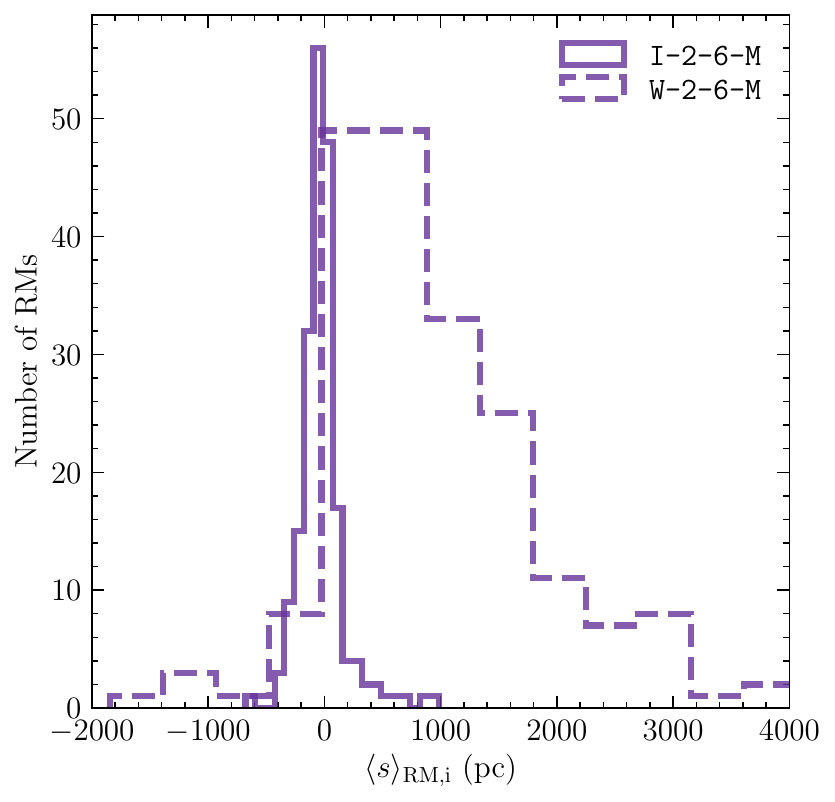} 
    \caption{Histogram of RMs in isolated (solid) and wind (dashed) runs with respect to RM-weighted mean distance from the LMC plane. The distribution is symmetric around the plane in the isolated case, while it is skewed towards the observer in the wind case.\\}
    \label{fig:RM_histogram}
\end{figure}

We show the (projected) mean positions defined by $\langle s \rangle_{\mathrm{RM},i}$ for each sightline as blue and red points in \autoref{fig:B_edgeplot}. These are the same points shown for the mock observations in \autoref{fig:RM_plot} with blue and red colours for positive and negative RMs, respectively, and the size corresponding to the RM magnitude.\footnote{Note that, in the projection shown in the plot, the LMC's line of nodes is tilted with respect to the normal to the plane of the figure. This tilt is the reason that the degree of separation between the positive RM (blue) and negative RM (red) sides of the galaxy is less visible here than it is, for example, in \autoref{fig:RM_plot}.}
We see that the presence of wind causes a significant shift in the location of maximum RM contribution along the LOS towards the observer. A considerable fraction ($>40\%$) of sightlines have $\langle s\rangle_{\mathrm{RM},i}>1$ kpc, while no sightlines have $\langle s\rangle_{\mathrm{RM},i}>1$ kpc in the run without the wind. In the presence of wind along the observer LOS, the electron density decays much more slowly than in its absence, leading to an order of magnitude higher $n_\mathrm{e}$ at 1 kpc from the plane in the wind run, with correspondingly stronger magnetic fields. The combined effects of increasing electron density and magnetic field strength in the tailwind cause the mean location of the RM signature to shift towards the observer. We can see that this shift is a general phenomenon in \autoref{tab:RM_table}, which reports the values of $\langle s\rangle_\mathrm{RM}$ and $\sigma_{s,\mathrm{RM}}$ for all our simulations, defined mathematically as
\begin{align}
\langle s \rangle_{\rm RM} &= \frac{\sum_i \int s n_\mathrm{e} |\vec{B}\cdot \hat{s}| \, ds}{\sum_i \int n_\mathrm{e} |\vec{B}\cdot \hat{s}| \, ds}, \label{eq:zRM} \\
\sigma_{s, \mathrm{RM}}^2 &= \frac{1}{N_\mathrm{LOS}} \sum_i\langle s \rangle_{\mathrm{RM, i}}^2, \label{eq:sigmaz}
\end{align}
where the sums run over our $N_\mathrm{LOS} = 200$ sightlines.

\autoref{fig:RM_histogram} shows a histogram of $\langle s \rangle_{\rm RM, i}$ values from \texttt{I-2-6-M} and \texttt{W-2-6-M} simulations, representing isolated and wind cases, respectively. Generally, for isolated cases, $\langle s \rangle_{\rm RM, i} \approx 0$, $\sigma_{s,\mathrm{RM}}< 200$ pc, and \autoref{fig:RM_histogram} shows that the distribution of $\langle s \rangle_{\rm RM, i}$ is narrowly-peaked near and symmetric about $s=0$ in the isolated case \texttt{I-2-6-M}. This means that RMs in the absence of a wind are entirely set by material close to the galactic plane, and that the signal comes equally from above and below the plane (as would be expected, since there is nothing to break the symmetry). By contrast, runs including the Milky Way CGM wind have $\langle s \rangle_{\rm RM}$ from $\sim 0.2-1$ kpc, $\sigma_{s,\mathrm{RM}}\sim 1$ kpc, and distributions of $\langle s\rangle_{\mathrm{RM},i}$ that are highly skewed to positive values as can be seen for the example of run \texttt{W-2-6-M} from \autoref{fig:RM_histogram}. This implies that when an observer looks at rotation measure from the LMC, the dominant signature arises from the tailwind on the near side rather than from the plane of the disc as has been conjectured by \citetalias{Gaensler2005} and \citet{Livingston2024}. Since the alignment fractions do not change significantly between the isolated and the wind cases, it also means that the magnetic field structure present in the disc is retained well out into the halo ($\sim 1$ kpc).

\begin{table}
\centering
\caption{Mean and dispersion of RM-weighted mean distance from the LMC plane.}
\begin{tabular}{ccc}
\hline\hline
\\[-2ex]
Simulation name \rev{(1)} & $\langle s \rangle_{\rm RM}$ \rev{(2)} & $\sigma_{s,\mathrm{RM}}$ \rev{(3)} \\ & \rev{(kpc)} & \rev{(kpc)}
\\[0.1ex]
\hline
\\[-2ex]
\texttt{I-0-2-L} & $\phantom{+}0.03$ & $0.17$\\
\texttt{I-2-0-L} & $\phantom{+}0.03$ & $0.09$\\
\texttt{I-2-2-L} & $\phantom{+}0.03$ & $0.13$\\
\texttt{I-2-6-L} & $\phantom{+}0.02$ & $0.16$\\
\texttt{W-2-0-L} & $\phantom{+}0.22$ & $0.70$\\
\texttt{W-2-2-L} & $\phantom{+}0.44$ & $0.88$\\
\texttt{W-2-6-L} & $\phantom{+}0.49$ & $0.94$\\
\texttt{I-0-12-M} & $-0.01$ & $0.17$\\
\texttt{I-2-6-M} & $-0.03$ & $0.19$\\
\texttt{W-2-6-M} & $\phantom{+}0.94$ & $1.41$\\
\texttt{I-2-6-H} & $-0.01$ & $0.17$\\
\\ \hline\hline
\end{tabular}\\[1.5ex]
\footnotesize{Columns: (1)\rev{: Simulation name --} name of simulation; (2)\rev{: $\langle s \rangle_{\rm RM}$ --} mean RM-weighted mean distance along LOS from the galactic plane over all sightlines; (3)\rev{: $\sigma_{s,\mathrm{RM}}$ --} dispersion of RM-weighted mean distances over all sightlines.
\\\vspace{0.1in}
}
\label{tab:RM_table}
\end{table}

\begin{figure}
    \includegraphics[width=1\columnwidth]{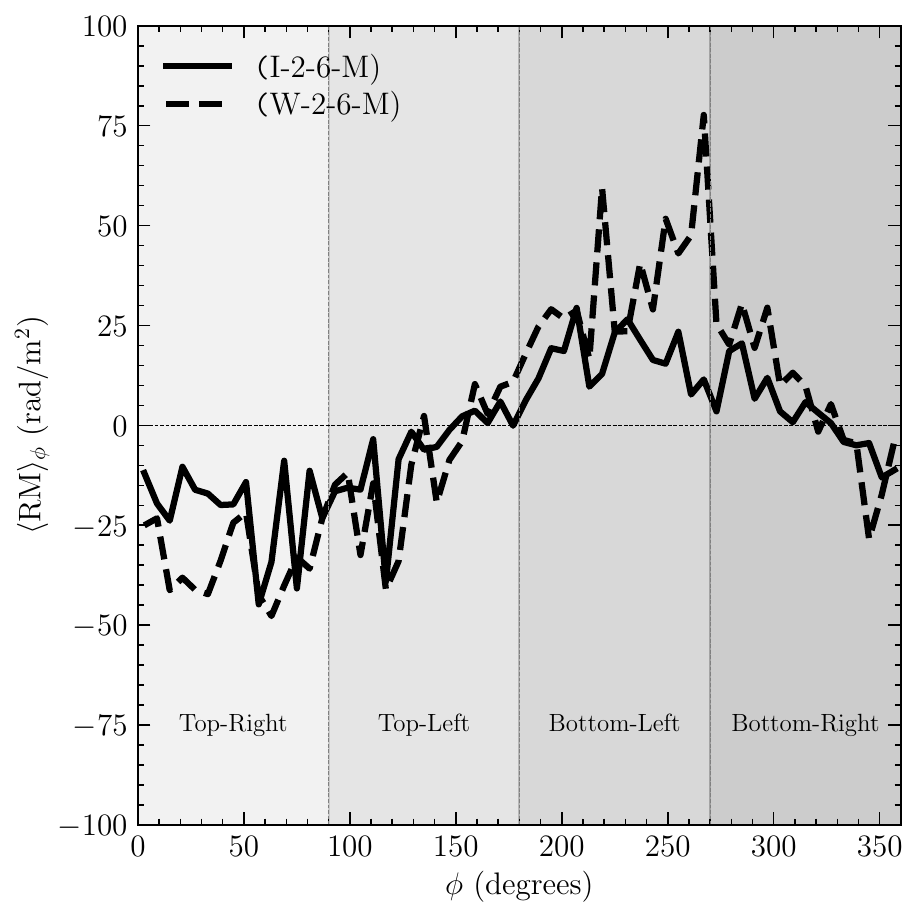} 
    \caption{\rev{Azimuthally averaged RM in bins of $6^{\circ}$ as a function of the azimuthal angle $\phi$. The four shaded regions show the quadrants in the mock RM map from \autoref{fig:RM_plot}. The enhancement in RM in the bottom-left (leading edge) is potentially a result of compression due to ram pressure stripping, consistent with the findings of \citet{bustard2020} and \citet{Livingston2024}. }\\}
    \label{fig:azimuthal_RM_average}
\end{figure}

\rev{\citet{bustard2020} simulations and \citet{Livingston2024} observations do predict one major signature of the Milky Way wind in the RM map: an enhancement in the field strength, and the associated RM, on the leading edge of the LMC where it encounters the wind. To investigate whether this signature is also present in our simulations, we compute mean RM as a function of azimuthal angle in \autoref{fig:RM_plot}, where we define $\phi = 0$ as parallel to the $x$ axis in the frame of the figure, with $\phi$ increasing counter-clockwise. We show the result in \autoref{fig:azimuthal_RM_average} for both runs \texttt{W-2-6-M} and \texttt{I-2-6-M}; the former is the run that best reproduces the RM statistics, and the latter is a control run that has identical initial conditions for the LMC but lacks the MW CGM wind. In the reference frame of the figure, the leading edge of the LMC lies in the third quadrant, $\phi = 180^\circ - 270^\circ$, and \autoref{fig:azimuthal_RM_average} shows that we indeed see a significant enhancement in the RMs there in run \texttt{W-2-6-M} compared to the control case \texttt{I-2-6-M} with no wind. Our results are therefore consistent with those of \citet{bustard2020} in this regard.}

\subsection{Which gas phase is responsible for the RM?}
\label{sec:GasPhase}

\begin{figure*}
    \includegraphics[width=2\columnwidth]{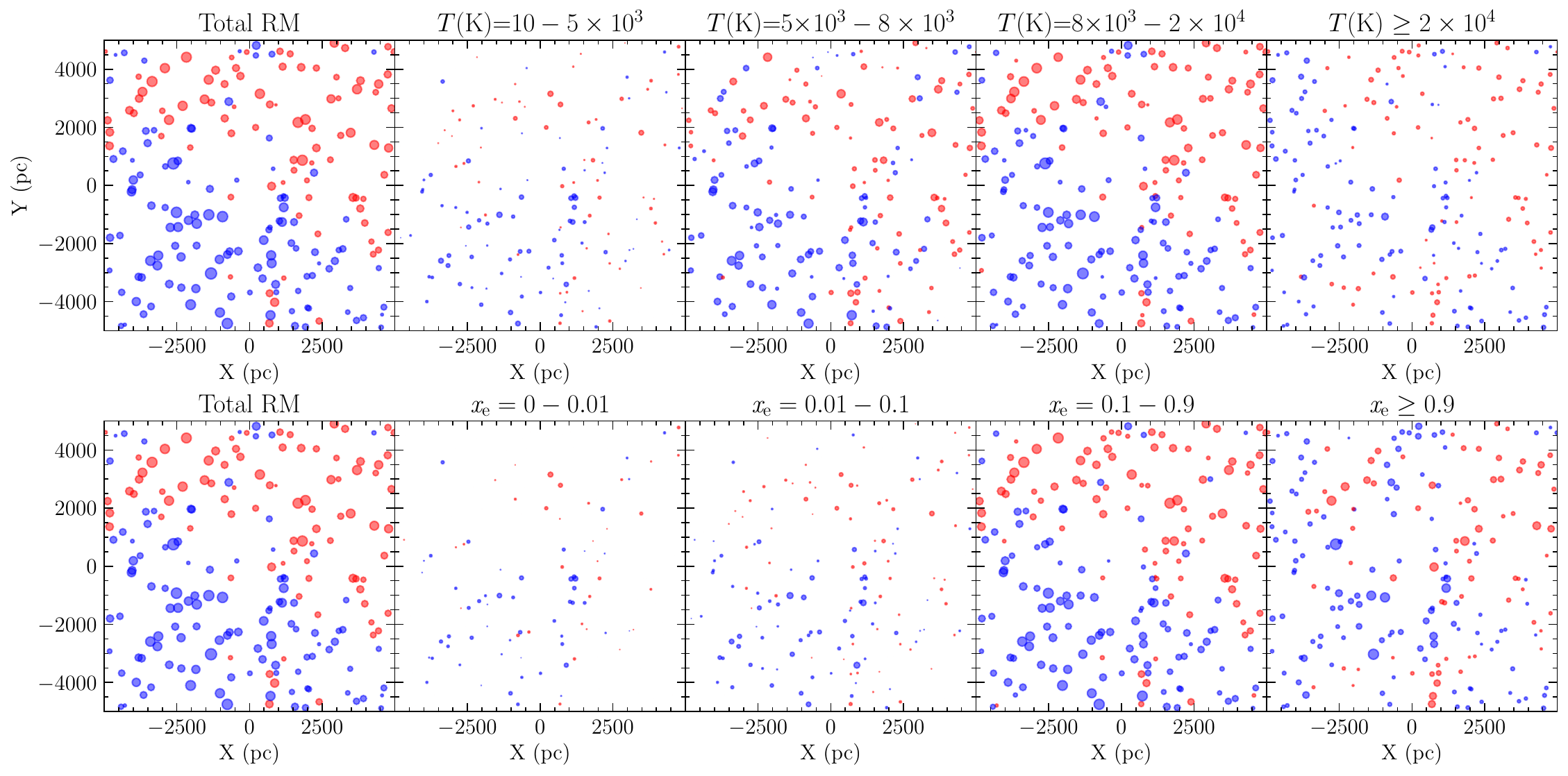} 
    \caption{RM maps due to gas with certain gas properties where blue (red) show positive (negative) RM values with the magnitude scaled to $\rm \propto |RM|^{1/2}$ and the largest circle corresponds to $\sim 130$ rad/m$^2$. The top row panels show gas categorisation based on its temperature. The leftmost panel is the overall RM map while the second, third, fourth, and fifth panels show RM maps from gas with temperatures between $10-5000$ K, $5000-8000$ K, $8000-20000$ K, and $\geq 20000$ K, respectively. The bottom row panels show gas categorisation based on its ionisation fractions ($x_\mathrm{e} = n_\mathrm{e}/n_\mathrm{H}$) where the leftmost panel is overall RM map again while the second, third, fourth, and fifth panels correspond to ionisation fractions between $0-0.01$, $0.01-0.1$, $0.1-0.9$, and $\geq 0.9$, respectively. These categories are carefully chosen to represent different ISM phases like the cold neutral medium, warm neutral medium, warm partially ionised medium, warm ionised medium, and fully ionised medium. In general, we find that the RM signature dominates from the warm ionised medium irradiated by the hard-UV photons from the extragalactic background. }
    \label{fig:RM_temperature_ion_analysis}
\end{figure*}

Now that we have identified where along the LOS the RM signature arises, we can further identify which gas phase is responsible for the RM signature. For this analysis, we again use simulation \texttt{W-2-6-M}, which best matches the observations. For this purpose we will divide the gas into different phases -- below we consider defining phases by temperature and ionisation fraction, and show that they all yield qualitatively-similar conclusions -- and generate RM maps for each phase where we manually set the RM contribution of all other phases to zero\rev{. This methodology} lets us isolate the contribution of each phase to the total RM. We complement these maps with statistics to characterise the fractional contribution of each phase; we define \rev{the} first of these statistics as
\begin{equation}
        f_{\mathrm{RM}, i, k} = \frac{|\mathrm{RM}|_{i,k}}{\sum_{k}|\mathrm{RM}|_{i,k}},
        \label{eq:fRM}
\end{equation}
where the first index, $i$, indicates a particular LOS, and second index represents a particular phase of interest\rev{. The} quantity in the denominator is the total RM along a given LOS. Thus\rev{,} \autoref{eq:fRM} quantifies the fractional contribution of some particular gas phase $k$ to the total RM signal measured along some sightline $i$. We further define
\begin{equation}
    \langle f_{\mathrm{RM},k} \rangle = \frac{\sum_i f_{\mathrm{RM},i,k} |\mathrm{RM}_i|}{\sum_i |\mathrm{RM}_i|},
    \label{eq:fRMtable}
\end{equation}
as the average contribution of each phase to the RM over all sightlines $i$, weighted by the RM magnitudes of all phases along each LOS, $|\mathrm{RM}_i|$. Our final statistic is the ``same sign fraction'' $f_\mathrm{RM-same}$, which we define as the fraction of sightlines for which the sign of the RM produced by a given phase matches the sign of the net RM over all phases along that sightline. In formal mathematical terms, we define this quantity as
\begin{equation}
    f_{\mathrm{RM-same},k} = \frac{1}{N_\mathrm{LOS}} \sum_i \Theta\left(\mathrm{RM}_{i,k} \, \mathrm{RM}_i\right),
    \label{eq:fRMsame}
\end{equation}
where $\mathrm{RM}_i$ is the total RM along sightline $i$ summing over all phases, and $\Theta(x)$ is the Heaviside step function, which returns 1 when the argument is positive and zero when it is negative. Thus, a low value of $f_{\mathrm{RM-same},k}$ implies that the RM from gas in phase $k$ is opposite in sign to the overall RM on a large fraction of sightlines.

\begin{table*}
\centering
\caption{Mean RM contribution and same sign fractions by gas phase.}
\begin{tabular}{cccccc}
\hline\hline
\\[0.1ex]
Temperature (K) & All $T$ & $10\leq T<5000$ & $5000 \leq T<8000$ & $8000 \leq T<20000$ & $T \geq 20000$ \\
[0.1ex]\\
\hline
\\[0ex]
$\langle f_{\mathrm{RM},k} \rangle~\%$ & $96.69$ & $1.9$ & $28.21$ & $64.85$ & $5.04$ \\
[0.7ex]
$f_{\mathrm{RM-same},k}$ \% & $100.0$ & $56.0$ & $75.0$ & $93.5$ & $75.5$\\
\\[-3ex]
\\ \hline\hline
\\[0.1ex]
$x_\mathrm{{e}}$ ($n_{{\rm e}}/n_{{\rm H}}$) & All $x_\mathrm{{e}}$ & $0\leq x_\mathrm{{e}}<0.01$ & $0.01\leq x_\mathrm{{e}}<0.1$ & $0.1 \leq x_\mathrm{e}<0.9$ & $x_\mathrm{e} \geq 0.9$\\
[0.1ex]\\
\hline
\\[0ex]
$\langle f_{\mathrm{RM},k} \rangle~\%$ & $96.41$ & $0.8$ & $1.83$ & $72.81$ & $24.57$\\
[0.7ex]
$f_{\mathrm{RM-same},k}$ \% & $100.0$ & $33.5$ & $56.5$ & $93.5$ & $74.5$\\
\\[-3ex]
\\ \hline\hline
\end{tabular}\\[1.5ex]
\footnotesize{The top part of the table shows gas divided into phases by temperature $T$, the bottom by electron abundance $x_e = n_e/n_\mathrm{H}$. The first row in each section reports the fractional contribution $\langle f_{\mathrm{RM},k}\rangle$ for a given phase averaged over all sightlines (\autoref{eq:fRMtable}), and the second row shows the same-sign fraction $f_{\mathrm{RM-same},k}$ (\autoref{eq:fRMsame}), which indicates the fraction of sightlines for which the sign of the RM for that phase matches the sign of the total RM. Columns show, from left to right, these quantities summed over all phases, and then for gas that lies in the indicated range of $T$ or $x_e$. Note that fractional contribution $\langle f_{\mathrm{RM},k}\rangle$ over all phases need not be exactly unity, because $\langle f_{\mathrm{RM},k}\rangle$ is normalised by the total RM, which is only equal to the sum of the individual phase RMs if all phases have the same sign (c.f.~\autoref{eq:fRM}).
\\\vspace{0.1in}
}
\label{tab:RM_frac_table}
\end{table*}

We show RM maps generated using two different ways of dividing up gas phases in \autoref{fig:RM_temperature_ion_analysis}; the maps in this figure follow the same convention as in previous RM maps, with blue (red) indicating positive (negative) RM values, and with the sizes of the circles scaled to $\rm \propto|RM|^{1/2}$. We also report the corresponding summary statistics for these phase divisions in \autoref{tab:RM_frac_table}.

In the top row of the figure, the gas is categorised based on its temperature. The leftmost panel shows the overall RM without any categorisation, and is thus identical to the RM plot shown in the middle panel of \autoref{fig:RM_plot}. The remaining panels show, from left to right, RM maps from gas with temperatures $10-5000$ K, $5000-8000$ K, $8000-20000$ K, and $\geq 20000$ K. We choose these temperature ranges to roughly divide the ISM into cold and unstable neutral medium (CNM and UNM), warm neutral medium (WNM), warm ionised medium (WIM), and hot ionised medium (HIM), respectively \citep{Begelman1990, Heiles2003, Kanekar2003, Cox2005, Roy2013, Murray2015, Murray2018, Bhattacharjee2023}. 

From both visual inspection and their low $\langle f_{{\rm RM},k}\rangle$ and $f_{\mathrm{RM-same},k}$ values (\autoref{tab:RM_frac_table}), it is clear that the CNM / UNM and HIM temperature ranges contribute negligibly to the overall RM. This is expected: in the cold medium the ionisation fraction is low, leading to a low electron density, while in the HIM the gas is fully ionised but the overall density is very low. By contrast, the WNM temperature regime contributes non-negligibly, $\langle f_{{\rm RM},k}\rangle\approx 28\%$. Gas in this temperature range is too cool to have significant thermal ionisation, but is assumed to form part of the Reynolds layer in our treatment (c.f.~\autoref{sec:PostProcess} and \autoref{fig:phase_plot}), and thus to be irradiated by the harder extragalactic background radiation field rather than the softer ISRF. This leads to non-thermal ionisation levels high enough to produce a notable RM contribution. However, the majority of the RM, $f_{\rm RM}\approx 65\%$, comes from WIM gas with temperatures in the range $8000-20000$ K. This temperature regime sees high ionisation levels due to the hard-UV radiation, photoionisation due to massive stars, and, at the very upper end of the temperature range, collisional ionisation.

The bottom row categorises gas based on its electron fraction $x_e=n_\mathrm{e}/n_\mathrm{H}$, where $n_\mathrm{H}$ is the number density of H nuclei\rev{. This} quantity ranges from 0 to $\approx 1.2$, with the latter case corresponding to full ionisation of both H and He. As in the top row, the leftmost panel shows the total RM over all phases, and in the remaining panels we break the gas into bins, from left to right, with $x_e$ \rev{=} $0-0.01$ (mostly CNM and UNM), $0.01-0.1$ (mostly WNM), $0.1-0.9$ (warm gas that is partially ionised by the extragalactic background rather than strong local sources), and $>0.9$ (WIM produced by strong stellar ionisation, together with thermally-ionised hot gas). We see that the dominant RM signature comes from partially-ionised gas with $x_e=0.1 - 0.9$; this phase has $\langle f_{{\rm RM},k}\rangle\approx73\%$, and $f_{\mathrm{RM-same},k}\approx 94\%$. The fully ionised bin, $x_e > 0.9$, is the second-largest contributor, and the remaining weakly-ionised phases are negligible. This result indicates that the bulk of the RM signature arises from gas where temperatures are not high enough for full ionisation, nor low enough to be part of the neutral medium, but instead exist in a regime of partial ionisation driven by the hard-UV photons of the extragalactic background rather than from softer local stellar sources (which produce $x_e \geq 1$). This points to an important result that the RM signature, at least for the LMC, arises from the Reynolds layer.

\section{Conclusions}
\label{sec:conclusions}
In this study, we perform \gizmo-MHD simulations of an LMC-analogue galaxy in two distinct environments - isolated and with ram pressure stripping due to wind experienced as a result of its motion through the density varying MW CGM. We consider a range of field magnetic strengths and configurations, and carry out detailed post-processing of our simulation results -- calibrated against DM data from pulsars -- in order to determine what magnetic conditions best reproduce  RM observations from the present-day LMC. Below, we summarise the main results.

\begin{enumerate}
    \item We succeed in identifying a configuration that shows very good agreement with the statistical properties of the observed LMC RM map; in particular, our simulation is able to reproduce the observed interquartile range, dispersion, and fraction of RM points which have a sign opposite \rev{of what} would be expected for a purely azimuthal field. In this simulation, the star-forming disc of the present-day LMC is characterised by a field that can reasonably be described as the sum of an ordered azimuthal component and a randomly-oriented turbulent component, with each component having a strength of $\approx 2$ $\mu$G.

    \item The simulation that best reproduces the observations includes the effects of the LMC's interaction with the MW CGM, which in the frame comoving with the LMC manifests as a wind that ram pressure-strips the galaxy. We compare our simulation including this wind to a simulation of an otherwise-identical but isolated LMC to deduce the effect of the wind on the galaxy's physical properties and RM statistics. We find that the wind produces a substantial, $\sim50\%$, enhancement in the azimuthal and the radial polar components of the magnetic field, and that this, in turn, leads to a $\sim 25\%$ increase in the magnitude of dispersion of RM. Only a minority of this enhancement is due to ram pressure compression; instead, it primarily arises because the magnetised MW CGM is deposited onto the LMC ISM as the LMC moves through the MW halo.

    \item The presence of wind also substantially modifies the physical location of the region responsible for producing the bulk of the RM signal. In the isolated case RMs probe a region very close to the LMC plane\rev{. But} in the interacting case the region that drives the RM is shifted towards the observer by $\sim 1$ kpc, so that it lies in a region dominated by a tail of highly magnetised, warm ($T\sim 10^4$ K), partially ionised (electron abundance $x_e \sim 0.1 - 0.9$) material stripped from the LMC by the MW wind\rev{. The partial ionisation occurs due to} exposure to the hard extragalactic radiation field that exists outside the LMC's Reynolds layer. This implies that RM measurements from background \rev{sources like AGNs} through the LMC are probing magnetic fields in the tailwind of the LMC rather than the galactic plane as previously thought (e.g., \citetalias{Gaensler2005}, \citealt{Livingston2024}).

\end{enumerate}

This study represents a new approach to the problem of deriving accurate magnetic field strengths and geometries from RM observations: forward modelling via simulations and mock observations rather than backward modelling that attempts to reconstruct field data from projected quantities. This new approach will become increasingly important as newer, higher-resolution RM data arrive from ongoing radio surveys such as ASKAP POSSUM \citep{Anderson21a, Vanderwoude24a}. It is clear from our work that analysing these data using traditional backward modelling techniques may lead to significant misinterpretations as a result of the assumptions one is inevitably forced to adopt with this method. Forward modelling as we have performed here offers us the opportunity for more robust results.

That said, we would be remiss not to mention some caveats to our results that we plan to address in future work. One clear one is the absence in our simulations both of the SMC and of a prominent stellar bar in the LMC (which is likely at least in part due to tidal forcing by the SMC). These features can potentially alter the gas dynamics and thus the magnetic field configuration, and including an SMC will obviously become necessary if we seek to model RM measurements of the Magellanic Bridge as well as in the body of the LMC. And as always in numerical work, resolution is a concern; while our mass resolution of 250 M$_\odot$ in the simulation that best reproduces the data is extremely high by the standards of most cosmological or even isolated galaxy simulations, it is not sufficient to allow us to forego the use of subgrid models to handle supernova and photoionisation feedback, and offers only marginal resolution of the dense molecular phase of the ISM. While it is unknown to what extent these resolution limitations influence magnetic structure, future simulations should strive for higher resolution to allay any concerns. We will pursue such higher-resolution models in future work.

\section*{Acknowledgements}
\rev{We thank the anonymous referee for the valuable comments that greatly improved the quality of this work. We thank Bryan Gaensler for providing \citetalias{Gaensler2005} data.} We thank Zipeng Hu for providing the code to generate synthetic LOS integrated quantities from \gizmo~ data. We would like to thank Marijke Havekorn, Katia Ferriere, Amit Seta, Roland Crocker, Hiep Nguyen, Neco Kriel, Aditi Vijayan, Antoine Marchal, Freeke van de Voort, Deovrat Prasad, Karlie Noon, along with other members from the groups of MRK and NMc-G for valuable discussions that greatly improved the quality of this study.

This research was partially funded by the Australian Government through Australian Research Council Australian Laureate Fellowships (project number FL220100020 awarded to MRK and project number FL210100039 awarded to NMc-G). This research was undertaken with the assistance of resources from the National Computational Infrastructure (NCI Australia), an NCRIS enabled capability supported by the Australian Government, through award jh2.

\section*{Data Availability}

Data related to this work will be shared upon reasonable request to the
corresponding author.

\bibliographystyle{mnras}

\bibliography{main}

\begin{appendix}

\section{Bootstrapping procedure}
\label{ap:boot}

This appendix outlines our bootstrapping method for analysing RM data from both observations and simulations, ensuring that we can compute statistics (e.g., outcome statistics mentioned in \autoref{tab:master_table}) on RM data sets properly marginalising over uncertainties, and do so in a manner that is consistent between real and mock observations.

Our first step when processing mock data is to generate realistic observational errors. To do so we first construct a kernel density estimate (KDE) of the distribution of $1\sigma$ uncertainties using the set of uncertainties reported by \citetalias{Gaensler2005}; we use \textsc{SciPy}\footnote{\url{https://docs.scipy.org/doc/scipy/reference/generated/scipy.stats.gaussian_kde.html}} for this purpose. For each realisation (see below), we then draw a value from that KDE for each sightline in our mock RM map. At this point our mock and observed maps both consist of a series of $N_\mathrm{LOS}$ sightlines for which we have both a central estimate of the rotation measure and an associated uncertainty $\{\mathrm{RM}_i, \sigma_i\}$.

Our next task is, given data in this format, to derive central estimates and uncertainties for any statistic $f(\{\mathrm{RM}_i\})$ that can be computed on a set of RM measurements; for example, we can take the function $f$ to be the variance function if we wish to compute the variance of RM, or the median function if we want to compute the median. Our bootstrapping method is the same regardless of the statistic we wish to evaluate, and it proceeds as follows. At each iteration, we generate a new realisation of the RM map in two steps:
\begin{enumerate}
    \item Generate a new set of $N_\mathrm{LOS}$ $\{\mathrm{RM}_i,\sigma_i\}$ pairs by drawing from the original sample \textit{with replacement}.
    \item For this new set of pairs, generate a new set of RM measurements $\{\mathrm{RM}'_i\}$ by, for each LOS produced in the previous step, drawing from a Gaussian distribution with mean $\mathrm{RM}_i$ and dispersion $\sigma_i$.
\end{enumerate}
This first step accounts for uncertainties induced by the finite number of sightlines to which we have access, while the second accounts for the uncertainties induced by the measurement errors on each sightline. We then calculate our target statistic on this realisation of the RM map as $f(\{\mathrm{RM}'_i\})$.

We repeat the above process $N$ times, yielding $N$ values of the statistic that we denote $f(\{\mathrm{RM}'_i\})_n$. We then compute the mean and standard deviation of these $N$ values, $\langle f(\{\mathrm{RM}'_i\})_n\rangle_N$ and $\mathrm{std}(f(\{\mathrm{RM}'_i\})_n)_N$. In order to ensure that the number of iterations $N$ is large enough to yield an accurate estimate of the mean and standard deviation, we proceed by sample doubling. We first carry out this procedure with $N=32$ trials, then repeat it for $N=64$ trials, and compare the values of $\langle f(\{\mathrm{RM}'_i\})_n\rangle_{32}$ and $\langle f(\{\mathrm{RM}'_i\})_n\rangle_{64}$. If these differ by less than 1\% we consider the procedure converged and accept $\langle f(\{\mathrm{RM}'_i\})_n\rangle_{64}$ and $\mathrm{std}(f(\{\mathrm{RM}'_i\})_n)_{64}$ as our final estimates of the mean and variance of the statistic $f$. If they differ by more than 1\%, we generate a sample of $N=128$ iterations and check if $\langle f(\{\mathrm{RM}'_i\})_n\rangle_{64}$ and $\langle f(\{\mathrm{RM}'_i\})_n\rangle_{128}$ differ by less than 1\%, and so forth, repeating the doubling procedure until two successive iterations differ by less than our 1\% tolerance, at which point we accept the largest $N$ values as our final estimates for the mean and standard deviation. In practice we find that most of our statistics reach convergence within 64-512 iterations.

For the simulated data, we must also marginalise over the first step of assigning error values to lines of sight by drawing from the KDE of observed errors. We do so by creating 100 different error realisations, i.e., 100 different $\{\mathrm{RM}_i, \sigma_i\}$ data sets. We apply the full bootstrapping procedure to each realisation and average the resulting statistics to produce our final values. For observations this step is not necessary, since we have only a single uncertainty $\sigma$ to go with each RM, and we therefore simply apply the bootstrapping process once with the actual measured errors.

\end{appendix}
\end{document}